\documentclass[11pt,aps,pra,notitlepage,nofootinbib,superscriptaddress,noeprint]{revtex4-2}
\usepackage[utf8]{inputenc}
\usepackage[english]{babel}
\usepackage[T1]{fontenc}
\usepackage{amsmath,amsfonts,amssymb,amsthm,bm,bbm,bbold}
\usepackage{graphicx}
\graphicspath{{Figures/}{./}}
\usepackage{mathtools}
\usepackage{physics}
\usepackage{makecell}
\usepackage{textcomp}
\usepackage{microtype}
\usepackage[dvipsnames]{xcolor}
\usepackage{dsfont}
\usepackage{booktabs}
\usepackage{ragged2e}
\usepackage[breaklinks=true,colorlinks=true,linkcolor=teal,urlcolor=teal,citecolor=teal]{hyperref}
\usepackage{orcidlink}

\begin{document}

\title{Charging of a Quantum Battery by a Two-Photon Quantum Pulse}

\author{Elnaz Darsheshdar\orcidlink{0000-0001-6341-7151}}
\thanks{darsheshdare@gmail.com}
\affiliation{Institute for Physical Research, Armenian National Academy of Sciences, Ashtarak-2, 0203, Armenia}

\author{Seyed Mostafa Moniri\orcidlink{0000-0003-1738-4429}}
\thanks{s.m.moniri@gmail.com}
\affiliation{Basic Sciences Group, Golpayegan College of Engineering, Isfahan University of Technology, Golpayegan 87717-67498, Iran}

\author{Mikayel Khanbekyan\orcidlink{0009-0009-2026-1164}}
\thanks{khanbekyan@gmail.com}
\affiliation{Institute for Physical Research, Armenian National Academy of Sciences, Ashtarak-2, 0203, Armenia}

\justifying

\begin{abstract}
\justifying
We investigate the charging of a harmonic-oscillator quantum battery by a propagating two-photon quantum pulse coupled through a two-level-system charger. Excitation-number conservation reduces the dynamics to a sequential response of the first and second excitation sectors, whose effective non-Hermitian generators exhibit two exceptional points separating overdamped, mixed, and underdamped regimes. We derive the exact full-charging amplitude and the response-matched two-photon temporal mode that achieves perfect charging in the ideal resonant single-channel model. For experimentally accessible Gaussian pulses, moderate temporal anticorrelation or a finite photon delay can enhance charging, whereas positive correlations generally suppress it. States with equal Schmidt number can nevertheless show different charging efficiencies, demonstrating that temporal-mode structure and response matching, rather than nonseparability alone, determine charging performance.
\end{abstract}


\maketitle

\section{Introduction}
\label{sec:introduction}

Quantum batteries represent a burgeoning class of devices designed to store and deliver energy using quantum mechanical resources such as coherence and entanglement~\cite{AlickiFannes2013,Binder2015,Campaioli2017,Ferraro2018,CampaioliRMP2024}. While many charging protocols rely on continuous classical driving or stationary energy sources~\cite{DowningUkhtaryPLA2024,DowningUkhtary2025}, few-photon quantum light provides a finite, controllable, and purely quantum energy resource for charging microscopic quantum systems~\cite{DarsheshdarMoniri2026}. In contrast with continuous classical driving, a two-photon pulse contains both a fixed excitation budget and a nontrivial joint temporal mode. 
This constraint motivates a fundamental question: how do photon arrival-time correlations and delays control the sequential loading of multiple energy quanta into a quantum battery? 

Temporal-mode engineering of few-photon states has also been studied in cavity-assisted single-emitter systems, including the generation of single-photon wave packets with controlled temporal profiles~\cite{Khanbekyan2008Pulse,Khanbekyan2017Controlled} and time-bin-entangled photon pairs from a three-level emitter~\cite{Khanbekyan2018TimeBin}. Related studies of pulsed single-photon spectroscopy have shown that the temporal and spectral structure of the incident photon, including chirp, internal vibrational dynamics, and mode-resolved descriptions, can strongly affect the information carried by the scattered field and the optimal measurement strategy~\cite{Albarelli2023,Darsheshdar2024Chirp,Khan2024QST,Khan2025Tensor,Das2025Vibrational}. These developments motivate treating the full joint temporal mode of the incident photon pair as a dynamical resource for quantum-battery charging.

Complementary continuous-variable two-photon charging mechanisms were studied by Downing and Ukhtary~\cite{DowningUkhtaryPLA2024,DowningUkhtary2025}. In Ref.~\cite{DowningUkhtaryPLA2024}, a harmonic-oscillator battery is driven directly by a quadratic field with a Gaussian pulse envelope, generating squeezing and an exponential population of higher even-number Fock sectors. In Ref.~\cite{DowningUkhtary2025}, this idea is embedded in a broader charger--battery setting, where a laser-driven bosonic charger transfers energy to the battery through linear or nonlinear coupling and the analysis emphasizes stored energy, charging power, ergotropy, and quadrature squeezing. Because these protocols are continuously driven, the available energy is not restricted to two quanta. By contrast, the present protocol is powered by a normalized propagating input state containing exactly two photons. It therefore addresses a distinct finite-resource and state-selective question: how efficiently can those two incident quanta be transferred into the prescribed battery state $|2\rangle_B$, and how can their joint temporal mode improve this transfer?

We consider a minimal charger--battery architecture consisting of a TLS coherently coupled to a harmonic oscillator. This is the two-photon extension of our previously solved single-photon charging model \cite{DarsheshdarMoniri2026}, in which the maximum stored energy, charging power, optimal incident mode, and exceptional-point charging time were obtained analytically. The TLS acts as a mediating charger that absorbs the incident pulse and transfers the resulting excitation to the battery. The central question is whether the two photons should arrive simultaneously or whether their joint temporal structure should be matched to the internal response of the charger--battery system.

The main results are as follows. First, the exact two-photon charging amplitude can be written as sequential convolutions of two internal response functions, associated with the first and second battery excitations. The corresponding effective non-Hermitian amplitude generators exhibit two exceptional points that separate overdamped, mixed, and underdamped dynamical regimes. Second, we identify the response-matched two-photon temporal mode that maximizes the full-charging probability at a prescribed target time. This mode achieves perfect charging in the ideal resonant single-channel model, whereas uncontrolled radiative loss imposes an upper bound on the charging probability. Third, we demonstrate that simultaneous photon arrival is generally suboptimal. Instead, moderate temporal anticorrelation or a finite pulse delay can significantly enhance the full-charging probability by matching the photon separation to the internal charger--battery response time. Finally, we reveal that the mean-squared deviation from the fully charged state depends nontrivially on the temporal correlation and delay parameters and directly quantifies the remaining charging imperfection.

Lengthy propagator and output-sector derivations are collected in the appendices.

\section{Theoretical Modeling}\label{sec:theory}
\begin{figure}[t]
	\centering
	\includegraphics[width=0.8\columnwidth]{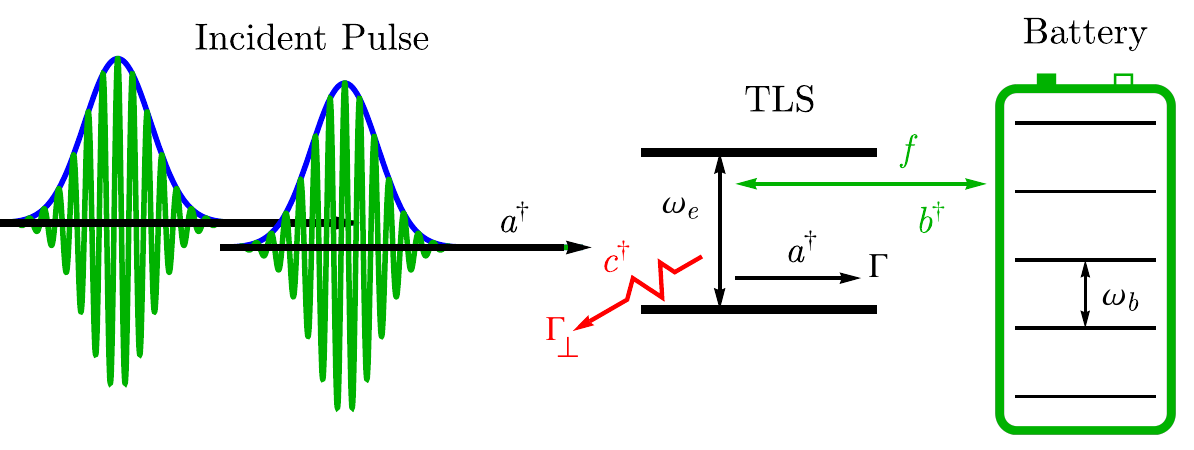}
	\caption{Schematic of the two-photon quantum-battery charging Protocol. A propagating two-photon wave packet couples to a two-level-system charger with rate $\Gamma$, while the charger is coherently coupled to the harmonic-oscillator battery with strength $f$. An additional decay channel with rate $\Gamma_{\perp}$ accounts for emission into uncontrolled electromagnetic modes.}
	\label{fig:scheme}
\end{figure}

We consider a traveling two-photon wave-packet interacting with a two-level system (TLS), where the TLS transfers the excitation by interacting with an isolated Harmonic oscillator (HO) serving as the battery, Fig.~\ref{fig:scheme}. The total Hamiltonian is written as $H=H_0+H_{\text{int}}$, with a single free part
\begin{equation}
	H_0=\omega_e\,\sigma^+\sigma^-+\omega_b\,b^\dagger b
	+\!\int_{0}^\infty d\omega\,\omega\,a^\dagger(\omega)a(\omega)
	+\!\int_{0}^\infty d\omega\,\omega\,c^\dagger(\omega)c(\omega),
\end{equation}
where $\omega_e$ is the transition frequency of the TLS between its ground state $\ket{g}$ and excited state $\ket{e}$, $ \omega_b$  is the frequency of the harmonic oscillator, $b$ and $b^\dagger$ are the annihilation and creation operators of the HO mode, satisfying $[b, b^\dagger] = 1$, and $\sigma^+\sigma^- = \ket{e}\bra{e}$. The traveling quantum light pulse is modeled as a continuum of bosonic modes with free Hamiltonian of $\int_0^\infty d\omega\, \omega\, a^\dagger(\omega) a(\omega)$, where $[a(\omega),a^\dagger(\omega')]=\delta(\omega-\omega')$. This finite photon-number wave-packet description is discussed in the general $N$-photon input-state framework of Ref.~\cite{Baragiola2012}. In addition to the pulse channel, the TLS may emit into uncontrolled electromagnetic environment that is described by a  bosonic reservoir with free Hamiltonian $\int_0^\infty d\omega\, \omega\, c^\dagger(\omega) c(\omega)$. Both the pulse and environment are coupled to the TLS through the standard dipole interaction $-\mathbf{d}\cdot\mathbf{E}$, 
where $\mathbf{d}$ is the TLS dipole operator and $\mathbf{E}$ is the electric field of the two channels.

The TLS is coherently coupled to the harmonic-oscillator battery through the exchange of single excitations. The corresponding interaction strength is denoted by $f$, which characterizes the coherent TLS--battery coupling.
We introduce the time-domain field operators
\begin{equation}
	a(t)=\frac{1}{\sqrt{2\pi}}\int d\omega\,
	a(\omega)e^{-i(\omega-\omega_e)t},
	\qquad
	c(t)=\frac{1}{\sqrt{2\pi}}\int d\omega\,
	c(\omega)e^{-i(\omega-\omega_e)t},
\end{equation}
which, within the white-noise approximation, satisfy $[a(t),a^\dagger(t')]=[c(t),c^\dagger(t')]=\delta(t-t')$. The temporal amplitudes used below are defined in the frame rotating at $\omega_e$, and the incident two-photon pulse is taken to be centered at the TLS transition frequency.
In the interaction picture with respect to $H_0$ and in the white-noise and rotating-wave approximations, we obtain the time-dependent interaction Hamiltonian of  
\begin{align}
	H^I(t) &= f\, \left( \sigma^+ b\, e^{i(\omega_e - \omega_b)t} + \sigma^- b^\dagger\, e^{-i(\omega_e - \omega_b)t} \right) \\ \nonumber
	&- i \left( \sqrt{\Gamma}  \sigma^+  a(t) - \text{h.c.} \right) - i \left( \sqrt{\Gamma_\perp}  \sigma^+  c(t) - \text{h.c.} \right).
\end{align}
 We consider the TLS and the harmonic-oscillator battery initially in their ground states, while the pulse field is prepared in a normalized two-photon wave-packet,
\begin{equation}
\label{init}
|\Psi(-\infty)\rangle=|g\rangle |0\rangle_B\otimes |2_{\phi_0}\rangle_P\otimes |0\rangle_E ,
\end{equation}
with
\begin{equation}
|2_{\phi_0}\rangle_P=\frac{1}{\sqrt{2}}\int_{-\infty}^{\infty} dt_1 dt_2\,\phi_0(t_1,t_2)a^\dagger(t_1)a^\dagger(t_2)|0\rangle_P .
\end{equation}
where $\phi_0$ is the normalized incoming two-photon wave-packet.
\begin{equation}
\phi_0(t_1,t_2)=\phi_0(t_2,t_1), \qquad \int dt_1dt_2\,|\phi_0(t_1,t_2)|^2=1 .
\end{equation}
In the following, we assume exact resonance between the TLS and the
harmonic-oscillator battery, $\omega_e=\omega_b$, and set the additional decay rate to zero, $\Gamma_\perp=0$, as both detuning and additional decay reduce the charging efficiency in the pulsed quantum-battery protocol considered in Ref.~\cite{DarsheshdarMoniri2026}.
Since the initial state of  Eq.~\eqref{init} contains two excitations and the interaction Hamiltonian conserves the total excitation number, the state remains in the two-excitation manifold in the form of 
\begin{equation}\label{ansatz2}
\begin{aligned}
|\Psi(t)\rangle=&\,\alpha_{e}(t)|e\rangle|1\rangle_B|0\rangle_P+\alpha_{g}(t)|g\rangle|2\rangle_B|0\rangle_P\\
&+\int d\tau\,\chi_{e}(t;\tau)|e\rangle|0\rangle_B a^\dagger(\tau)|0\rangle_P+\int d\tau\,\chi_{g}(t;\tau)|g\rangle|1\rangle_B a^\dagger(\tau)|0\rangle_P\\
&+\frac{1}{\sqrt{2}}\int d\tau_1 d\tau_2\,\phi(t;\tau_1,\tau_2)|g\rangle|0\rangle_B a^\dagger(\tau_1)a^\dagger(\tau_2)|0\rangle_P .
\end{aligned}
\end{equation}
where $\alpha_e(t)$, $\alpha_g(t)$, $\chi_e(t;\tau)$, $\chi_g(t;\tau)$,
and $\phi(t;\tau_1,\tau_2)$ are probability amplitudes. The two-photon amplitude is symmetric, $\phi(t;\tau_1,\tau_2)=\phi(t;\tau_2,\tau_1)$ with the initial condition $\phi(-\infty;\tau_1,\tau_2)=\phi_0(\tau_1,\tau_2)$.

We denote by $U(t)\equiv U(t,-\infty)$ the interaction-picture evolution operator generated by $H^I(t)$. Then, the normal-ordered form of the Schr{\"o}dinger equation reads
\begin{equation}
\begin{aligned}
\frac{d}{dt}|\Psi(t)\rangle=&\left[-if(\sigma^+b+\sigma^-b^\dagger)-\gamma \sigma^+\sigma^-\right]|\Psi(t)\rangle\\
&-\sqrt{\Gamma}\sigma^+U(t)a(t)|\Psi(-\infty)\rangle+\sqrt{\Gamma}a^\dagger(t)\sigma^-|\Psi(t)\rangle .
\end{aligned}
\end{equation}
where $\gamma={\Gamma}/{2}$.


In Appendix~\ref{app:time_dynamics} we show the exact solution of the time dynamics in this system. We reduce the two-photon input to a single-photon problem and then propagate the resulting amplitudes through the one- and two-excitation sectors.  Finally the exact two-photon charging amplitude of the state $|g\rangle|2\rangle_B|0\rangle_P$, is obtained as 
\begin{equation}
\alpha_g(t)=-2\Gamma f^2\int_{-\infty}^{t}ds\int_{-\infty}^{s}du\, G_2(t-s)G_1(s-u)\phi_0(u,s),
\label{alphag_final_sym2}
\end{equation}
which has a sequential convolution form of the response functions
$G_1$ and $G_2$.
The probability of full two-excitation charging i.e., $P_2(t)=|\alpha_g(t)|^2$ is given by:
\begin{equation}
P_2(t)=\left|-2\Gamma f^2\int_{-\infty}^{t}ds\int_{-\infty}^{s}du\,G_2(t-s)G_1(s-u)\phi_0(u,s)\right|^2.
\label{P2_final}
\end{equation}
Throughout this paper, we use the term "fully charged battery" to refer to the case in which both excitation quanta available in the two-photon input are stored in the battery, corresponding to the preparation of the target state $|2\rangle_B$.

\subsection{Dynamical regimes of the response functions}

The two-photon charging amplitude is governed by two internal response functions. The function $G_1(t)$ describes the response associated with the first battery excitation, while $G_2(t)$ describes the response in the two-excitation sector. The homogeneous amplitude dynamics in the two sectors can be written
in the common form
\begin{equation}
    M_j=
    \begin{pmatrix}
        -\gamma & -i\sqrt{j}\,f \\
        -i\sqrt{j}\,f & 0
    \end{pmatrix},
    \qquad j=1,2,
    \label{eq:Mj-general}
\end{equation}
where $j=1$ and $j=2$ correspond to $M_1$ and $M_2$, respectively. The eigenvalues of $M_j$ are
\begin{equation}
    \lambda_{\pm}^{(j)} = \frac{-\gamma \pm\sqrt{\gamma^2-4jf^2}}{2},
    \label{eq:Mj-eigenvalues}
\end{equation}
and, therefore, exceptional points of the effective internal dynamics obtained after eliminating the propagating continuum exist at $f_{\mathrm{EP}}^{(j)}  =  \gamma/(2\sqrt{j})$. Related exceptional-point structures in effective open-system descriptions have been investigated in cavity-QED and coupled-cavity settings, including their influence on spontaneous-emission dynamics and resonant response~\cite{KhanbekyanWiersig2020,Khanbekyan2023EP}.
For the single- and two-excitation sectors they occur at $f_{\rm EP}^{(1)}={\gamma}/{2}$ and $f_{\rm EP}^{(2)}={\gamma}/{\sqrt{8}}$, correspondingly. 
These two points divide the dynamics into three distinct dynamical regimes. For positive time arguments, the response functions can be written in the piecewise form
\begin{equation}
G_j(t)=
\begin{cases}
\displaystyle
e^{-\gamma t/2}\frac{\sinh(\Delta_j t)}{\Delta_j},&D_j>0,\\[12pt]
\displaystyle
t e^{-\gamma t/2},&D_j=0,\\[12pt]
\displaystyle
e^{-\gamma t/2}\frac{\sin(\Omega_j t)}{\Omega_j},&D_j<0,
\end{cases}
\label{Gj_piecewise}
\end{equation}
where
\begin{equation}
D_j=\gamma^2-4jf^2
\qquad
\Delta_j=\frac{1}{2}\sqrt{D_j},
\qquad
\Omega_j=\frac{1}{2}\sqrt{-D_j}.
\end{equation}
\begin{figure*}[t]
    \centering
    \begin{minipage}{0.32\textwidth}
    \centering
    \includegraphics[width=\linewidth]{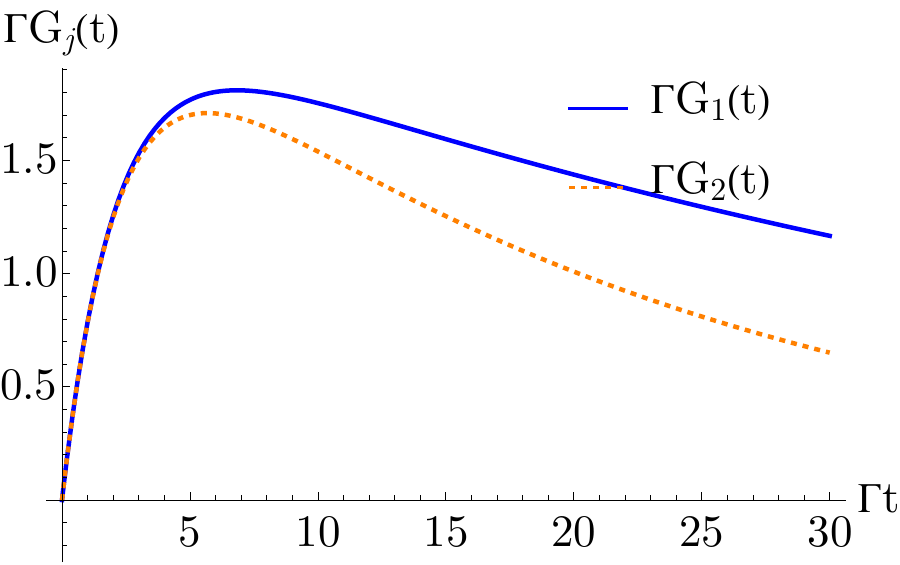}\\[-1mm]
    \textbf{(a)}
    \end{minipage}
    \hfill
    \begin{minipage}{0.32\textwidth}
    \centering
    \includegraphics[width=\linewidth]{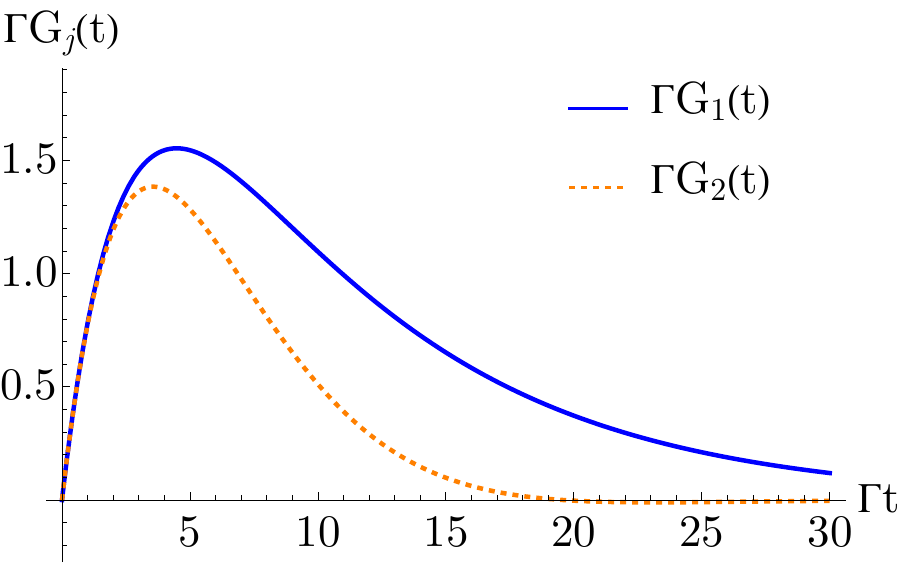}\\[-1mm]
    \textbf{(b)}
    \end{minipage}
    \hfill
    \begin{minipage}{0.32\textwidth}
    \centering
    \includegraphics[width=\linewidth]{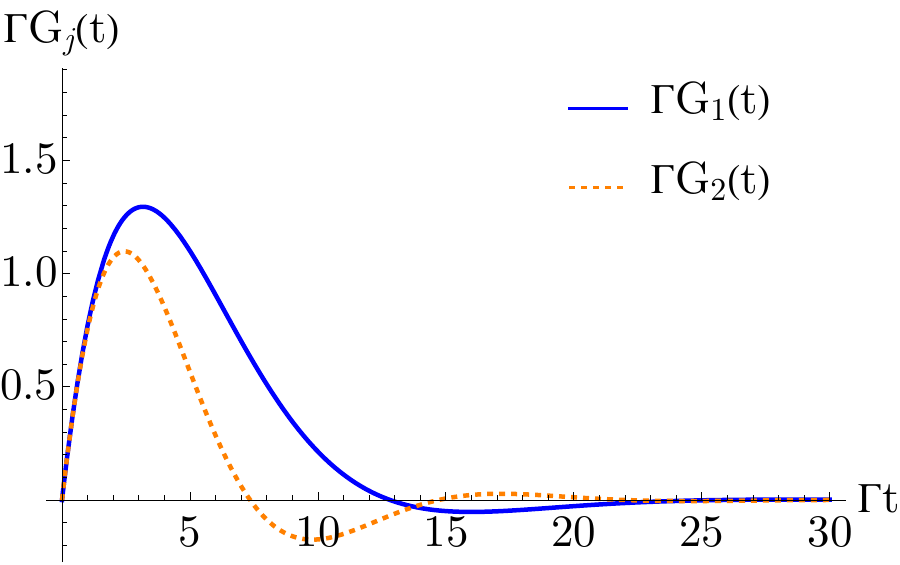}\\[-1mm]
    \textbf{(c)}
    \end{minipage}
    \caption{
    Single- and two-excitation response functions in the three dynamical regimes. The solid curves show $\Gamma G_1(t)$, governing the transfer of the first excitation, and the dashed curves show $\Gamma G_2(t)$, governing the transfer of the second excitation. The panels correspond to (a) $f/\Gamma=0.10$, below both effective exceptional points; (b)  $f/\Gamma=0.21$, between the two exceptional points, where $G_1$ remains nonoscillatory while $G_2$ is oscillatory; and (c) $f/\Gamma=0.35$, above both exceptional points, where both responses are oscillatory. Here $\gamma=\Gamma/2$, so that $f_{\mathrm{EP}}^{(2)}/\Gamma=1/(4\sqrt{2})\simeq0.1768$ and $f_{\mathrm{EP}}^{(1)}/\Gamma=1/4$.}
    \label{fig:response-regimes}
\end{figure*}
Figure ~\ref{fig:response-regimes} illustrates the qualitative change of the internal charging response across the two critical couplings. Below both exceptional points, $G_1$ and $G_2$ are positive, nonoscillatory functions with long dissipative tails. Between the exceptional points, the enhanced coupling $\sqrt{2}f$ in the two-excitation sector is already sufficient to produce an oscillatory response $G_2$, while $G_1$ remains overdamped. Above both exceptional points, both response functions oscillate and change sign. Since the charging kernel contains the product $G_2(t-s)G_1(s-u)$, these sign changes represent phase reversals of the charging amplitude and imply that temporal-mode phase matching, in addition to photon separation, becomes important in the strong-coupling regime.

For $f<{\gamma}/{\sqrt{8}}$, both response functions are non-oscillatory. The coherent TLS--battery coupling is not strong enough to overcome the effective decay of the excited TLS amplitude. As a result, the excitation transfer from the TLS to the battery occurs in a relaxational manner rather than through coherent exchange oscillations. The charging process is therefore dominated by irreversible decay and monotonic response functions. Physically, both the first and the second excitation transfer are limited by dissipation, so the battery is charged inefficiently unless the incoming two-photon wave packet is well matched to the slowly decaying response.

For ${\gamma}/{\sqrt{8}}<f<{\gamma}/{2}$, the single-excitation response $G_1(t)$ remains overdamped, while the two-excitation response $G_2(t)$ becomes oscillatory. This regime is specific to the two-photon charging dynamics. The first excitation is still transferred to the battery through a non-oscillatory, dissipative response, but once one excitation has been stored, the transition $|e\rangle|1\rangle_B\leftrightarrow|g\rangle|2\rangle_B$ is governed by the enhanced coupling $\sqrt{2}f$. Therefore the second excitation sector can already support coherent TLS--battery exchange even though the first excitation sector cannot. The charging dynamics in this regime is mixed: the first transfer acts as a dissipative bottleneck, while the second transfer can show oscillatory enhancement or suppression depending on the photon arrival-time structure.

For $f>{\gamma}/{2}$, both $G_1(t)$ and $G_2(t)$ are oscillatory. The coherent coupling between the TLS and the battery dominates over the effective dissipative decay, so the excitation can be exchanged reversibly between the charger and the battery before being emitted into the probe channel. The charging dynamics is therefore no longer monotonic. Instead, the battery population can display coherent peaks and revivals. In this regime the timing of the two photons becomes especially important, because the second photon can either arrive in phase with the internal exchange and enhance the full two-excitation charging probability, or arrive out of phase and suppress it.

\subsection{Analytical optimal two-photon wave packet}
\label{sec:optimal_two_photon}

In this sub-section, we show an analytical form of the optimal two-photon wave packet. Although the preparation of such a bi-variate pulse shape is experimentally challenging, we present it as a benchmark for various pulse shapes and to explain how experimentally achievable wave packets can charge the battery efficiently through their overlap with the optimal one. 

From Eq.~\eqref{alphag_final_sym2}, we can write the two-photon charging amplitude at a given target time $t_f$ in terms of the ordered charging kernel as
\begin{align}
\alpha_g(t_f)=-2\Gamma f^2
\int_{-\infty}^{t_f}ds
\int_{-\infty}^{s}du\,
K_{t_f}(u,s)\phi_0(u,s),
\label{eq:alpha_kernel}
\end{align}
where the ordered kernel is
\begin{align}
K_{t_f}(u,s)=G_2(t_f-s)G_1(s-u)\Theta(t_f-s)\Theta(s-u).
\label{eq:ordered_kernel}
\end{align}

As a result of the bosonic symmetry and normalization, the two ordered regions contribute equally, so the norm over $u<s$ is $1/2$ and using the Cauchy--Schwarz inequality on this ordered interval we have
\begin{align}
P_2(t_f)\le 2\Gamma^2f^4\,\mathcal N_2(t_f),
\label{eq:P2_bound}
\end{align}
where
\begin{align}
\mathcal N_2(t_f)=\int_{-\infty}^{t_f}ds
\int_{-\infty}^{s}du\,\left|G_2(t_f-s)G_1(s-u)\right|^2,
\label{eq:N2}
\end{align}
which can be evaluated analytically, as shown in Appendix~\ref{app:optimal_two_photon}. The obtained optimal full-charging probability is
\begin{equation}
P_{2,\mathrm{opt}}=\frac{\Gamma^2}{4\gamma^2}.
\label{eq:P2_ultimate}
\end{equation}
In the case of the system without uncontrolled radiative loss we have, $\gamma=\Gamma/2$, and therefore
\begin{equation}
P_{2,\mathrm{opt}}=1.
\label{eq:P2_unit}
\end{equation}
Thus this  two-photon wave-packet will charge the battery completely at a given target time of $t_f$. This result holds for any nonzero coupling $f$. In the limit $f\rightarrow0$ the response-matched temporal mode acquires an increasingly long duration, while for $f=0$ no excitation transfer to the battery is possible.

The symmetric wave packet of 
\begin{align}
\phi_{\rm opt}(t_1,t_2;t_f)=\frac{
\mathcal K_{t_f}^*(t_1,t_2)+\mathcal K_{t_f}^*(t_2,t_1)}{
\sqrt{2\mathcal N_2(t_f)}}
\label{eq:phi_opt_two}
\end{align}
with
\begin{align}
\mathcal K_{t_f}(t_1,t_2)=G_2(t_f-t_2)G_1(t_2-t_1)
\Theta(t_f-t_2)\Theta(t_2-t_1),
\label{eq:K_full}
\end{align}
will saturate the bound. The two terms of the Eq.~\eqref{eq:phi_opt_two} are related to the two time-ordered halves of the joint temporal area and reflect bosonic symmetry.
\begin{figure}[t]
    \centering
    \includegraphics[width=0.78\columnwidth]
    {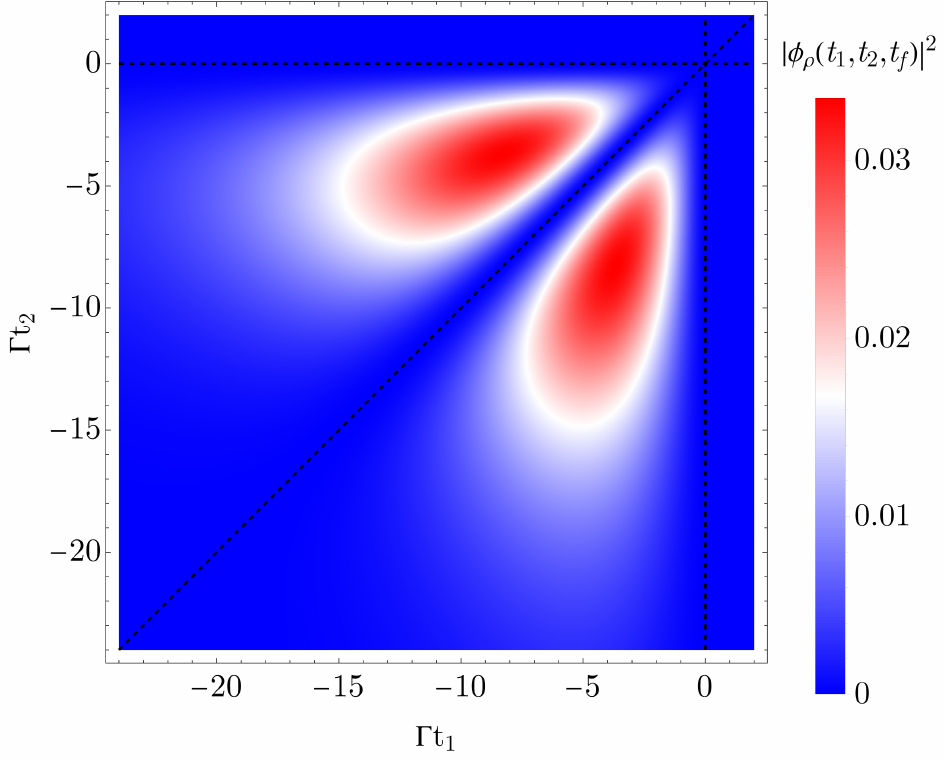}
    \caption{
    Optimal two-photon temporal probability density $|\phi_{\rm opt}(t_1,t_2;t_f)|^2$ for $f/\Gamma=0.20$ and $\Gamma t_f=0$. The wave packet is obtained directly from the response-matched product of the single- and two-excitation response functions, $G_1$ and $G_2$. The dashed diagonal denotes simultaneous photon arrival, $t_1=t_2$. Since $G_1(0)=0$, the optimal temporal mode vanishes along this line, demonstrating that simultaneous arrival is not optimal for the sequential two-excitation charging process. The vertical and horizontal dotted lines indicate the prescribed target time $t_f$.}
    \label{fig:optimal_two_photon_mode}
\end{figure}

In Figure~\ref{fig:optimal_two_photon_mode} we show the temporal structure of the optimal two-photon wave packet. The two symmetric areas are related to two possible time orderings of photons. They are nonzero only for times before the given charging time. In addition the probability density vanishes along the diagonal $t_1=t_2$ as we have $G_1(0)=0$ for simultaneous photon arrival. Thus, the optimal mode suppresses the arrival of both photons at the same time and favors a finite separation time between photons, i.e., a separation time comparable to the characteristic response time of $G_1$.

The optimal photon state is non-Gaussian. Its structure is directly related to the sequential charging process: the earlier photon is weighted by the first-excitation response $G_1$, while the later photon is weighted by the second-excitation response $G_2$. The correlated and delayed Gaussian states studied below should therefore be regarded as experimentally achievable pulses. 

\subsection{Effect of an environmental decay channel}
\label{sec:uncontrolled-loss}

In the presence of a non-zero environmental decay rate $\Gamma_{\perp}$, the excited-state amplitude decay rate is 
\begin{equation}
    \gamma_{\mathrm{tot}}=\frac{\Gamma+\Gamma_{\perp}}{2}.
\end{equation}
The two-photon charging amplitude retains the form of Eq.~\eqref{alphag_final_sym2}, with the response functions $G_1$ and $G_2$ evaluated using $\gamma_{\rm tot}$ instead of $\gamma$ , thus we have 
\begin{equation}
    \int_0^\infty
    dt\,|G_j(t)|^2=\frac{1}{2\gamma_{\mathrm{tot}}jf^2},
\end{equation}
with
\begin{equation}
    \mathcal N_2=\frac{1}{8\gamma_{\mathrm{tot}}^2f^4}.
\end{equation}
The optimal two-photon-charging probability is 
\begin{equation}
    P_{2,\mathrm{opt}}=\frac{\Gamma^2}{4\gamma_{\mathrm{tot}}^2}= \left(\frac{\Gamma}{\Gamma+\Gamma_{\perp}}\right)^2\equiv \beta^2 .
\end{equation}

We assign $\phi_{\mathrm{opt}}^{(\perp)}(t_1,t_2;t_f)$ being the optimal two-photon wave-packet obtained from Eq.~\eqref{eq:phi_opt_two} (replacing $\gamma$ with $\gamma_{\mathrm{tot}}=(\Gamma+\Gamma_{\perp})/2$ in the response functions of $G_1$ and $G_2$). For an arbitrary normalized input temporal mode and using Eq.~\eqref{alphag_final_sym2}, the charging probability can then be written as
\begin{equation}
P_2(t_f)=\beta^2\left|\left\langle\phi_{\mathrm{opt}}^{(\perp)}(t_f)\middle|\phi_0\right\rangle\right|^2\leq \beta^2 .
\end{equation}
This expression highlights two types of constraints on the charging
probability. First the factor $\beta^2$ shows the maximum probability which is allowed in the presence of the environmental loss.  Second the overlap of $\left|\langle\phi_{\mathrm{opt}}^{(\perp)}|\phi_0\rangle\right|^2$ that represents the compatibility of the temporal shape of the incident pulse with the optimal wave-packet.
As expected, even for perfect temporal mode matching, the charging probability cannot exceed $\beta^2$ for $\Gamma_{\perp}\neq0$. If the input pulse is not perfectly matched to the optimal mode, the charging probability is further reduced.

The quadratic dependence on $\beta$ reflects the sequential nature of the two-photon charging process: both excitation-transfer steps must proceed through the useful channel without emission into the uncontrolled reservoir.

In the presence of additional lose into the environments the  exceptional points  are  
\begin{equation}
f_{\mathrm{EP},\perp}^{(j)}=\frac{\Gamma+\Gamma_{\perp}}{4\sqrt{j}}, \qquad j=1,2.
\end{equation}

The exceptional points are shifted since the additional environmental loss increases the excited-state damping from $\gamma$ to $\gamma_{\mathrm{tot}}$. Therefore, the critical TLS--battery coupling strength is shifted to larger values.

\section{Gaussian two--photon pulses}
In this section, we use the experimentally achievable Gaussian wave-packet to investigate how the temporal shape of this incoming photon pair affects the probability of two-photon charging of the battery. Since the two-excitation charging is a sequential process, the relative arrival time of the two photons will play an important role on this procedure. We study two kinds of temporal-mode shaping: a correlated Gaussian two-photon wave-packet, defined by a correlation coefficient $\rho$, and a  delayed Gaussian wave packet, defined by a delay time of $\tau$. 

\subsection{Temporally entangled photons }
We first consider the correlated Gaussian wave packet
\begin{equation}
\varphi_{\rho}(t_1,t_2) = \frac{1}{\sqrt{2\pi T^2\sqrt{1-\rho^2}}} \exp\left[-\frac{t_1^2+t_2^2-2\rho t_1t_2}{4T^2(1-\rho^2)}\right], 
\label{correlatedGause}
\end{equation}
where $-1<\rho<1$. Positive $\rho$ is related to the case that photons arrive together, while a negative $\rho$ is related to different arrival times of each photon. Zero $\rho$ is for factorized two-photon pulse of 
\begin{equation}
\varphi_{\rm fac}(t_1,t_2)=\xi(t_1)\xi(t_2),
\end{equation}
with the single-photon pulse as a normalized Gaussian wave packet,
\begin{equation}
\xi(t)=\frac{1}{(2\pi T^2)^{1/4}} \exp\left(-\frac{t^2}{4T^2}\right),    
\end{equation}
where $T$ is the pulse duration. 

\begin{figure*}[t]
    \centering
    \includegraphics[width=\textwidth]
    {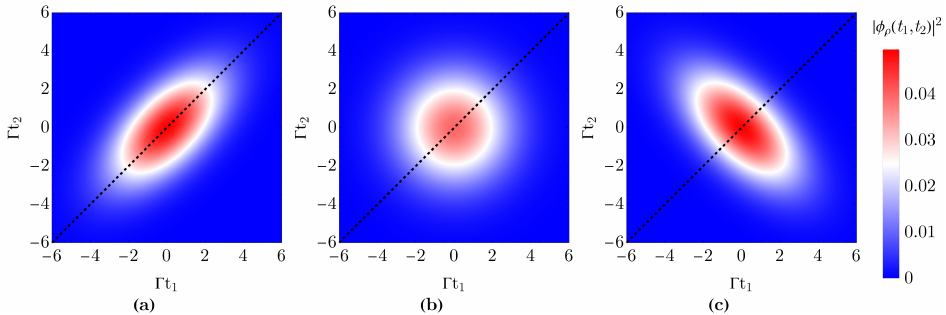}
    \caption{
    Joint temporal probability densities $|\phi_\rho(t_1,t_2)|^2$ for $\Gamma T=2$ and (a) $\rho=+0.6$, (b) $\rho=0$, and (c) $\rho=-0.6$. The dashed line denotes simultaneous arrival, $t_1=t_2$. Positive correlation concentrates the photon-pair distribution along this line, whereas negative correlation broadens the distribution in the relative-time direction. Although the states with $\rho=\pm0.6$ have the same effective Schmidt number, $K_\rho=1.25$, their orientations relative to the ordered charging kernel are different.}
    \label{fig:probabilities}
\end{figure*}

The correlated Gaussian amplitude of Eq. \ref{correlatedGause} is non-factorizable in the case of $\rho\neq0$. The Schmidt decomposition is a quantifier that we use to show this point \cite{Mikhailova2008,JeronimoMoreno2009}. We define
\begin{equation}
\rho_1(t,t')=\int_{-\infty}^{\infty}dt_2\,
\varphi_\rho(t,t_2)\varphi_\rho^*(t',t_2).
\label{eq:reduced_temporal_state}
\end{equation}
Thus a direct evaluation gives
\begin{equation}
\mathcal{P}_\rho=\operatorname{Tr}\!\left(\rho_1^2\right)= \sqrt{1-\rho^2},\qquad
K_\rho=\frac{1}{\mathcal{P}_\rho}=\frac{1}{\sqrt{1-\rho^2}},
\label{eq:schmidt_rho}
\end{equation}
where $K_\rho$ is the  Schmidt number. The factorized pulse of $\rho=0$ gives $K_\rho=1$, and $K_\rho>1$ for all nonzero $\rho$. The value of $K_\rho$ depends only on $|\rho|$; thus, the states with coefficients $+\rho$ and $-\rho$ have the same degree of temporal-mode correlation. In contrast, their temporal profiles are oriented differently, as shown in Fig.~\ref{fig:probabilities}. Positive correlation directs the two-photon temporal distribution around the line of $t_1\simeq t_2$ that favors both photons having nearly the same interaction time with TLS. In contrast negative correlation gathers the distribution along the $t_1\simeq- t_2$ direction and favors different interaction times for each photon. 
This point can also be understood using  the definition of the squared widths for the joint probability distribution along the difference and sum coordinates of
\begin{equation}
\sigma_-^2=2T^2(1-\rho),\qquad
\sigma_+^2=2T^2(1+\rho),
\end{equation}
where $\sigma_-^2$ and $\sigma_+^2$ are defined as the variances of $t_1-t_2$ and $t_1+t_2$, respectively. Negative correlation would broaden the distribution of relative interaction times of two photons with the TLS. It does not impose a finite delay and the center of the Gaussian distribution is still at $t_1=t_2=0$.

The orientation of the temporal distribution is important for the dynamics of the system since the charging amplitude is determined by contributions from the time-ordered domain $u<s<t$. That is why two states with coefficients $+\rho$ and $-\rho$ can have the same Schmidt number but substantially different charging efficiencies, as shown below. Within the two-photon family states we study here, the charging performance is controlled not by the amount of temporal-mode entanglement, but instead by the distribution of the joint temporal correlations with respect to the sequential charger--battery response.

\subsection{Two time-delayed pulses }
In addition to correlated pulse discussed earlier, we consider a delayed Gaussian pulse (in symmetrized form for simplicity) as
\begin{equation}
\varphi_{\tau}(t_1,t_2)=\frac{\xi(t_1+\tau/2)\xi(t_2-\tau/2)+\xi(t_1-\tau/2)\xi(t_2+\tau/2)}{\sqrt{2+2\exp\left(-\tau^2/4T^2\right)}}.
\end{equation}
where $\tau$ is the time delay between two incoming pulses. In the limit of $\tau=0$ the two pulses are completely overlapped. We use a normalized form which is considered to compensate  the finite overlap between the two delayed temporal wave-packets. Both cases of entangled and delayed photons would be equal to the factorized Gaussian pulses when $\tau=\rho=0$ which have
\begin{equation}
\varphi_{\rho=0}(t_1,t_2)=\varphi_{\tau=0}(t_1,t_2)=\xi(t_1)\xi(t_2).
\end{equation}
We studied the maximum probability of two-excitation state of the HO battery,
\begin{equation}
P_2^{\max}=\max_t |\alpha_g(t)|^2,
\end{equation}
and compare it with the factorized case using enhancement ratios of
\begin{equation}
\eta_\rho=\frac{P_2^{\max}(\rho)}{P_2^{\max}(0)}, \qquad
\eta_\tau=\frac{P_2^{\max}(\tau)}{P_2^{\max}(0)}.
\end{equation}

\begin{figure}
\centering
\begin{minipage}{0.48\textwidth}
\centering
\includegraphics[width=\linewidth]{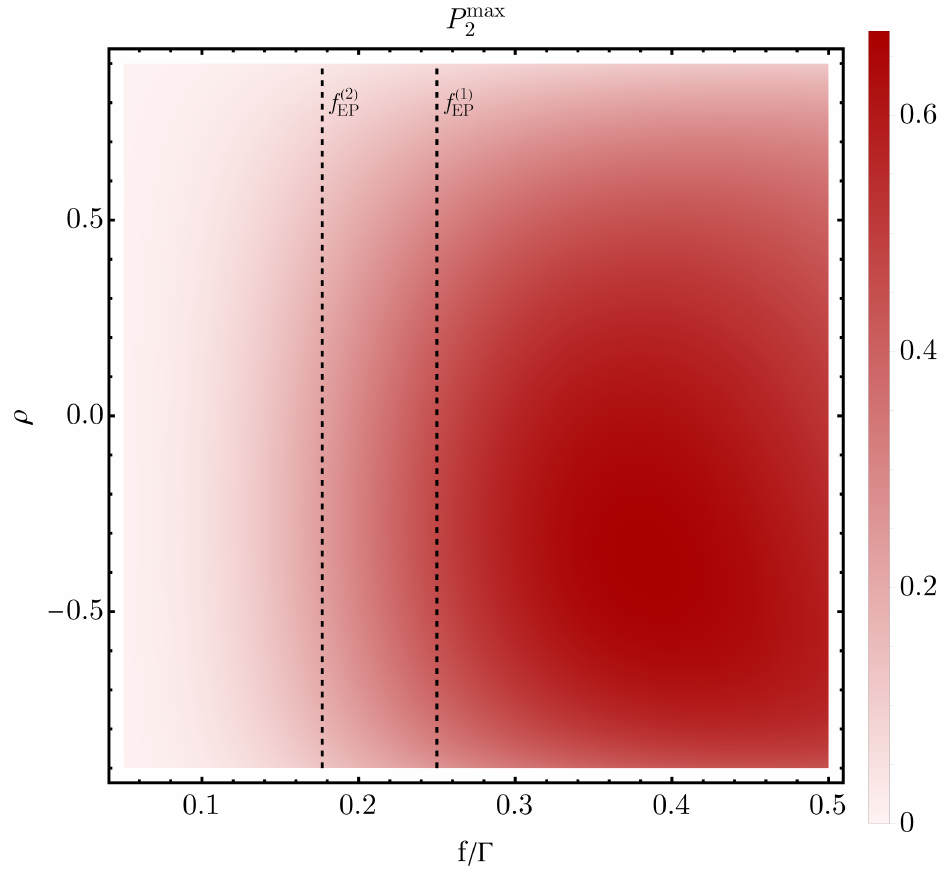}\\[-1mm]
\textbf{(a)}
\end{minipage}
\hfill
\begin{minipage}{0.48\textwidth}
\centering
\includegraphics[width=\linewidth]{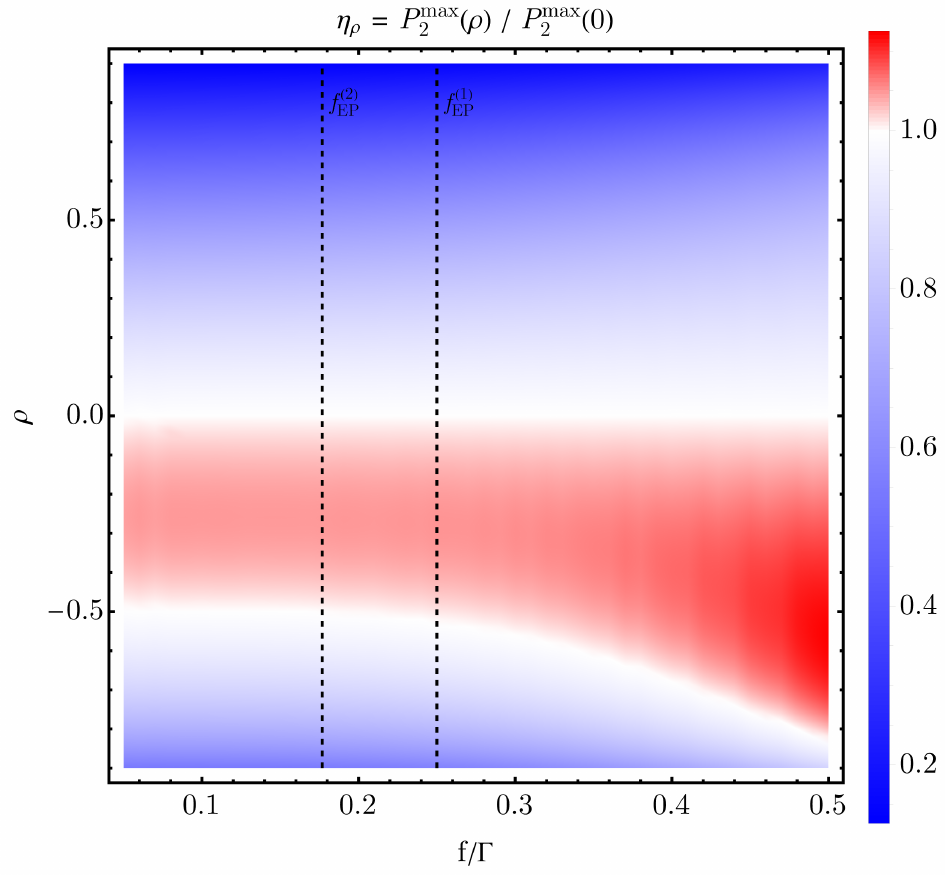}\\[-1mm]
\textbf{(b)}
\end{minipage}   
\caption{Correlated Gaussian two-photon pulse for $\Gamma T=2$. (a) Maximum two-excitation charging probability $P_2^{\max}$ as a function of the normalized TLS--battery coupling $f/\Gamma$	and the temporal-correlation coefficient $\rho$. (b) Relative enhancement $\eta_\rho=P_2^{\max}(\rho)/P_2^{\max}(0)$, where $\rho=0$ corresponds to the factorized Gaussian reference pulse. Positive $\rho$ describes photons that tend to arrive together, whereas negative $\rho$ favors temporally separated arrival times. The vertical dashed lines indicate the two dynamical exceptional points $f_{\rm EP}^{(2)}=\gamma/(2\sqrt{2})$ and $f_{\rm EP}^{(1)}=\gamma/2$. Moderate negative correlations enhance the charging probability relative to the uncorrelated reference, while positive correlations generally suppress it.}   
\label{fig:rho_gaussian}
\end{figure}

\begin{figure}[t]
\centering
\begin{minipage}{0.48\textwidth}
\centering
\includegraphics[width=\linewidth]{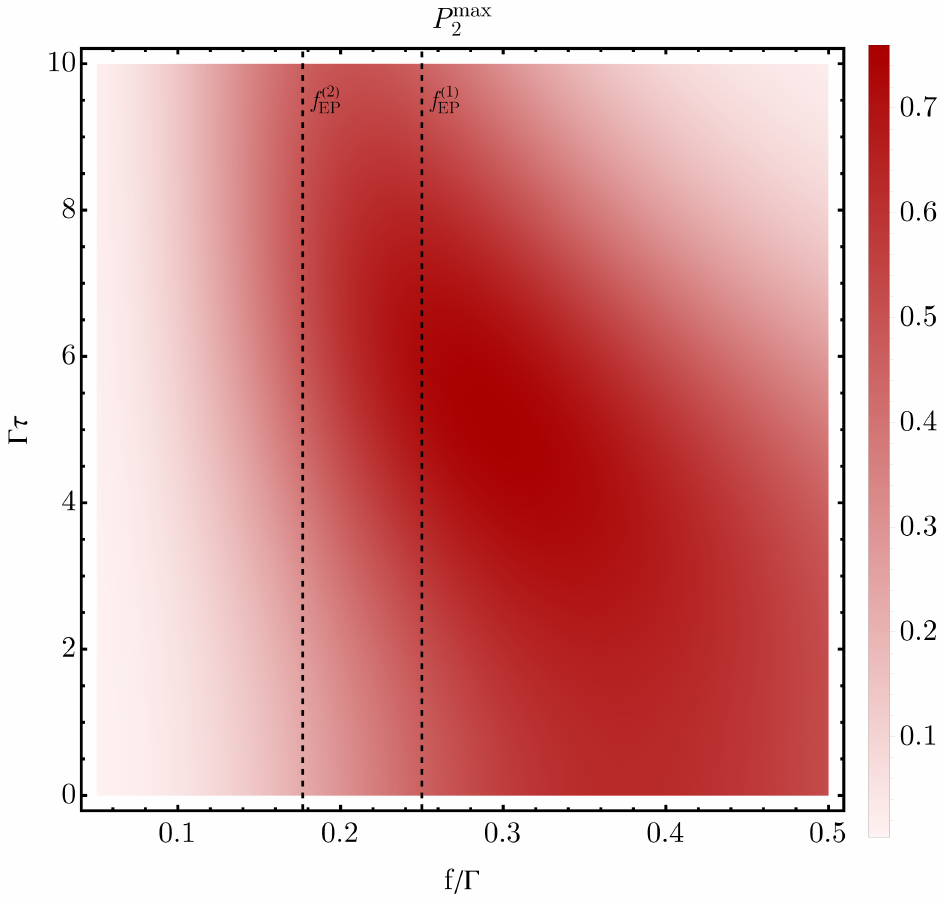}\\[-1mm]
\textbf{(a)}
\end{minipage}
\hfill
\begin{minipage}{0.48\textwidth}
\centering
\includegraphics[width=\linewidth]{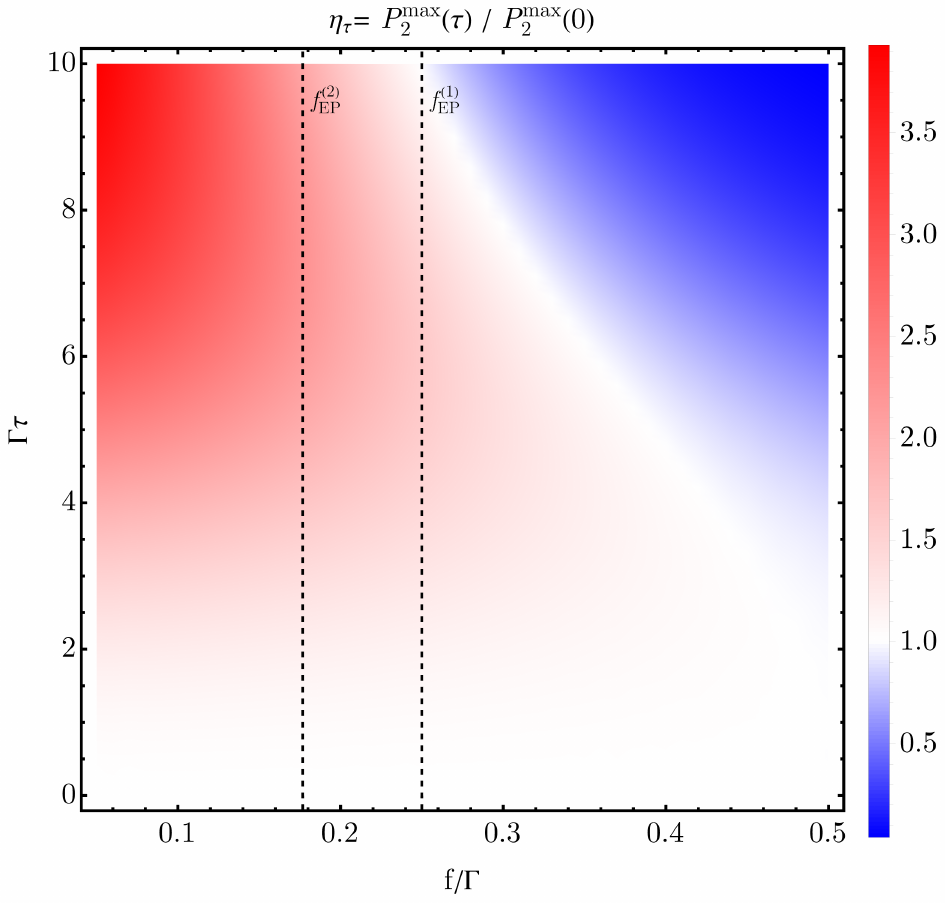}\\[-1mm]
\textbf{(b)}
\end{minipage}
\caption{
Symmetrized delayed Gaussian two-photon pulse for $\Gamma T=2$. (a) Maximum two-excitation charging probability $P_2^{\max}$ as a function of the normalized TLS--battery coupling $f/\Gamma$ and the dimensionless delay time $\Gamma\tau$. (b) Relative enhancement $\eta_\tau=P_2^{\max}(\tau)/P_2^{\max}(0)$, where $\tau=0$ corresponds to two temporally overlapping Gaussian modes and therefore to the same factorized Gaussian reference pulse. The vertical dashed lines indicate the two exceptional points $f_{\rm EP}^{(2)}=\gamma/(2\sqrt{2})$ and $f_{\rm EP}^{(1)}=\gamma/2$. A finite delay can enhance the charging probability by matching the arrival time of the second photon to the internal charger--battery response, while excessive delay leads to dynamical mismatch.}
\label{fig:delayed_gaussian}
\end{figure}

Figure ~\ref{fig:rho_gaussian} shows the absolute charging probability and the enhancement factor $\eta_\rho$ for the correlated Gaussian pulse. The corresponding results for the delayed pulse are shown in Fig.~\ref{fig:delayed_gaussian}. Both figures are calculated for the fixed pulse duration $\Gamma T=2$. Figure ~\ref{fig:rho_gaussian} reveals a clear asymmetry between positive and negative correlations. Positive correlations generally reduce the full-charging probability, whereas moderate negative correlations enhance it. Very strong negative correlations again become unfavorable because the typical photon separation exceeds the relevant internal response time.
Although states with $+|\rho|$ and $-|\rho|$ have the same Schmidt number $K_\rho$, their charging probabilities are different. Thus, the Schmidt number alone is insufficient to predict the charging performance. The sign and temporal structure of the correlations also play an essential role.

The result for the delayed Gaussian pulse in Fig. \ref{fig:delayed_gaussian} gives a more direct benchmark for the effect of different interaction times of each photon. For $\tau=0$, the two photons completely overlap in time, while increasing $\tau$ separates the two pulses. The map of $P_2^{\max}(f,\tau)$ shows that a finite delay can substantially improve the charging probability, especially when the delay becomes comparable to the characteristic timescale over which the first excitation is transferred through the TLS–battery system. For large delays, the dominant contribution to the sequential charging amplitude contains the factor $G_1(s-u)$ with $s-u\sim\tau$. Since the single-excitation response vanishes in the long-delay limit, $|G_1(\tau)|\to 0$ as $\tau\to\infty$, the contribution of the delayed two-photon pulse to the sequential charging amplitude becomes negligible. As a consequence, the two-photon charging probability is suppressed for very large delays. In the oscillatory regime, this decay is modulated by oscillations, while the overall envelope still decreases exponentially. This confirms that the simultaneous arrival of the photons is unfavorable for the two-photon charging process, but a temporal separation that is matched with the sequential transfer of the two excitations is useful.

From Fig. \ref{fig:delayed_gaussian}b one can see that the enhancement ratio $\eta_\tau$ is large in the weak-coupling limits. This should not be interpreted as the weak-coupling regime giving the best absolute charging performance, since Fig. \ref{fig:delayed_gaussian}a indicates that the probability $P_2^{\max}(\tau=0)$ is small there. In fact, it shows that simultaneous pulses are inefficient in this regime and introducing a delay yields a large relative improvement.

To compare the Gaussian pulse results directly with the exact response-matched solution, we optimize the correlation coefficient or the delay separately for each value of the TLS--battery coupling. Since $P_2^{\max}$ already denotes the maximum over the charging time, we define
\begin{equation}
	P_{2,\rho}^{\mathrm{best}}(f)=
	\max_{\rho} P_2^{\max}(f,\rho),
	\qquad
	P_{2,\tau}^{\mathrm{best}}(f)=
	\max_{\tau} P_2^{\max}(f,\tau).
\end{equation}
The factorized reference is denoted by $P_{2,\mathrm{fac}}^{\max}(f)$.

\begin{figure}[t]
	\centering
	\includegraphics[width=0.8\linewidth]{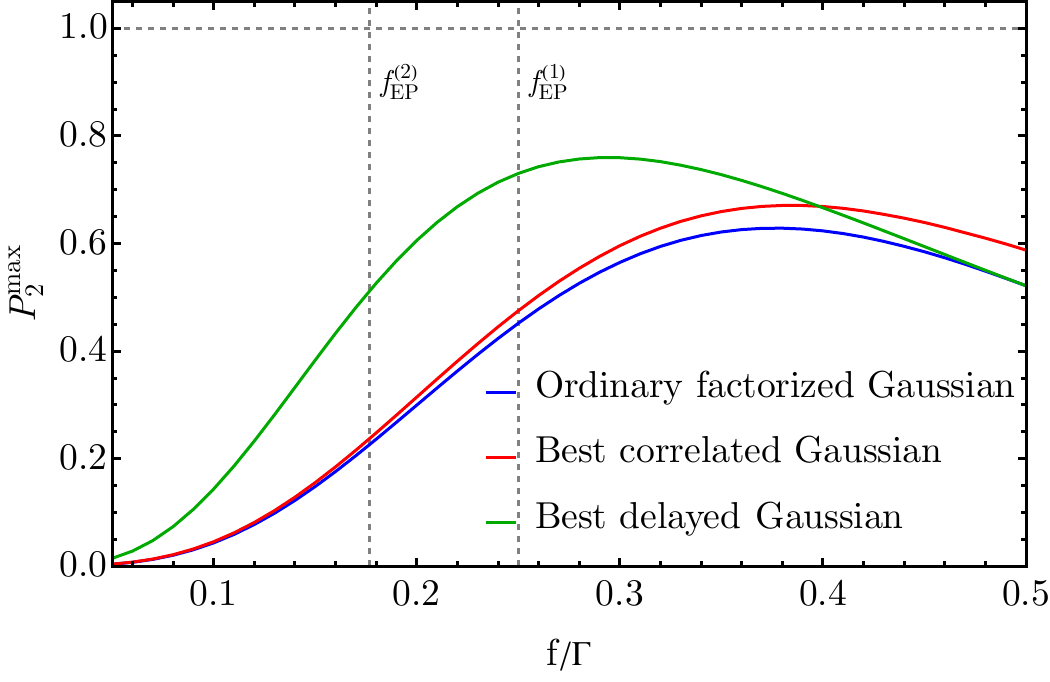}
	\caption{Maximum full-charging probability as a function of the	normalized TLS--battery coupling $f/\Gamma$ for Gaussian two-photon pulses with $\Gamma T=2$. The factorized Gaussian, $P_{2,\mathrm{fac}}^{\max}$, is compared with the optimized	correlated Gaussian, $P_{2,\rho}^{\mathrm{best}}$, and the optimized delayed Gaussian, $P_{2,\tau}^{\mathrm{best}}$. The horizontal line	shows the exact response-matched optimum, $P_{2,\mathrm{opt}}=1$, obtained when the full two-photon temporal mode is independently optimized at each nonzero coupling. The vertical lines indicate the two effective exceptional points $f_{\mathrm{EP}}^{(2)}$ and $f_{\mathrm{EP}}^{(1)}$.}
	\label{fig:benchmark}
\end{figure}
Figure~\ref{fig:benchmark} compares the performance of the restricted Gaussian pulse results with the exact response-matched limit. Optimization of the delay gives the largest improvement over the factorized Gaussian over most of the coupling range, while optimization of the temporal correlation produces a smaller enhancement. 

\subsection{Full charging probability and deviation from the fully charged state}
\label{sec:charging_precision}
The two-photon-charging probability $P_2(t)$ gives the probability of finding the battery in the two-excitation state. To characterize the distance from this target state directly, we resolve the battery population into the zero-, one-, and two-excitation sectors,
\begin{align}
P_2(t) &= |\alpha_g(t)|^2,\\
P_1(t) &= |\alpha_e(t)|^2+\int d\tau'\,|\chi_g(t;\tau')|^2,\\
P_0(t) &= 1-P_1(t)-P_2(t),
\end{align}
where $P_n(t)$ is the probability that the harmonic-oscillator battery contains $n$ excitations.

The average energy stored in the harmonic-oscillator battery is
\begin{equation}
E_B(t)=\omega_b\left[|\alpha_e(t)|^2+\int d\tau\,|\chi_g(t;\tau)|^2+2|\alpha_g(t)|^2\right].
\label{EB_def}
\end{equation}
The normalized stored energy, measured relative to the maximum two-quantum energy $2\omega_b$, is therefore
\begin{equation}
\eta_E(t)\equiv\frac{E_B(t)}{2\omega_b}=\frac{1}{2}\left[|\alpha_e(t)|^2+\int d\tau\,|\chi_g(t;\tau)|^2+2|\alpha_g(t)|^2\right].
\label{normalized_energy}
\end{equation}
This normalized energy has a direct interpretation in the present
finite-resource setting:
\begin{equation}
\eta_E(t)=\frac{P_1(t)+2P_2(t)}{2}.
\label{eq:energy_transfer_efficiency}
\end{equation}
The numerator, $P_1(t)+2P_2(t)$, is the mean number of excitations stored in the battery, whereas the denominator represents the two excitation quanta carried by the resonant two-photon input. Therefore, $\eta_E(t)$ gives the fraction of the total available input energy that is stored in the battery at time $t$, and $0\leq \eta_E(t)\leq 1$. In particular, $\eta_E(t)=1$ corresponds to storage of the full two-photon energy, while smaller values indicate that only part of the available energy has been transferred to the battery.

Because the incident field contains exactly two photons, the total available input energy is fixed to $2\omega_b$, and $\eta_E$ directly measures the fraction of this energy stored in the battery. The mean stored energy, however, does not fully characterize the charging quality. For example, population of the singly excited sector can increase $E_B$ while the desired full-charging probability $P_2$ remains below unity. To quantify how closely the battery approaches the target excitation number $n_B=2$, we therefore introduce the mean-squared deviation 
\begin{equation} 
	\mathcal D_2(t) = \left\langle\left(2-n_B\right)^2\right\rangle = 4P_0(t)+P_1(t), 
	\label{eq:mean_squared_deviation} 
	\end{equation} 
where $n_B=b^\dagger b$. Thus, $\eta_E$ characterizes the overall energy-transfer efficiency, whereas $P_2$ and $\mathcal D_2$ provide additional information on the preparation of the target state $|2\rangle_B$.

The battery-energy variance,
\begin{equation}
\operatorname{Var}(H_B)=\langle H_B^2\rangle-\langle H_B\rangle^2,
\end{equation}
could be an ambiguous measure of charging quality. In particular, it vanishes for each number state $|0\rangle_B$, $|1\rangle_B$, and $|2\rangle_B$, even though only the last one is fully charged.

On the other hand in $\mathcal D_2(t)=4P_0(t)+P_1(t)$, the weights assigned to the zero- and one-excitation sectors follow directly from their squared distances from the target excitation number $n_B=2$: $(2-0)^2=4$, $(2-1)^2=1$, and $(2-2)^2=0$. Consequently, $\mathcal D_2(t)\geq0$ and 
\begin{equation}
\mathcal D_2(t)=0
\quad\Longleftrightarrow\quad
P_2(t)=1.
\end{equation}
Thus, unlike the energy variance, $\mathcal D_2$ has a unique charging interpretation: the battery population is concentrated near the fully charged two-excitation state if $\mathcal D_2$ is small.

For each pulse-control parameter $\lambda$, we evaluate this quantity at the time at which the full-charging probability is maximal, 
\begin{equation}
t_2^*(\lambda)=\operatorname*{arg\,max}_t P_2(t;\lambda),
\qquad
P_2^{\max}(\lambda)=P_2\!\left[t_2^*(\lambda);\lambda\right],
\label{eq:tstar_definition}
\end{equation}
where $\lambda=\rho$ for the correlated Gaussian pulse and $\lambda=\tau$ for the delayed Gaussian pulse. We then define
\begin{equation}
\mathcal D_{2,*}(\lambda)=\mathcal D_2\!\left[t_2^*(\lambda);\lambda\right].
\label{eq:deviation_at_tstar}
\end{equation}

\begin{figure}[t]
\centering
\begin{minipage}{0.48\textwidth}
\centering
\includegraphics[width=\linewidth]{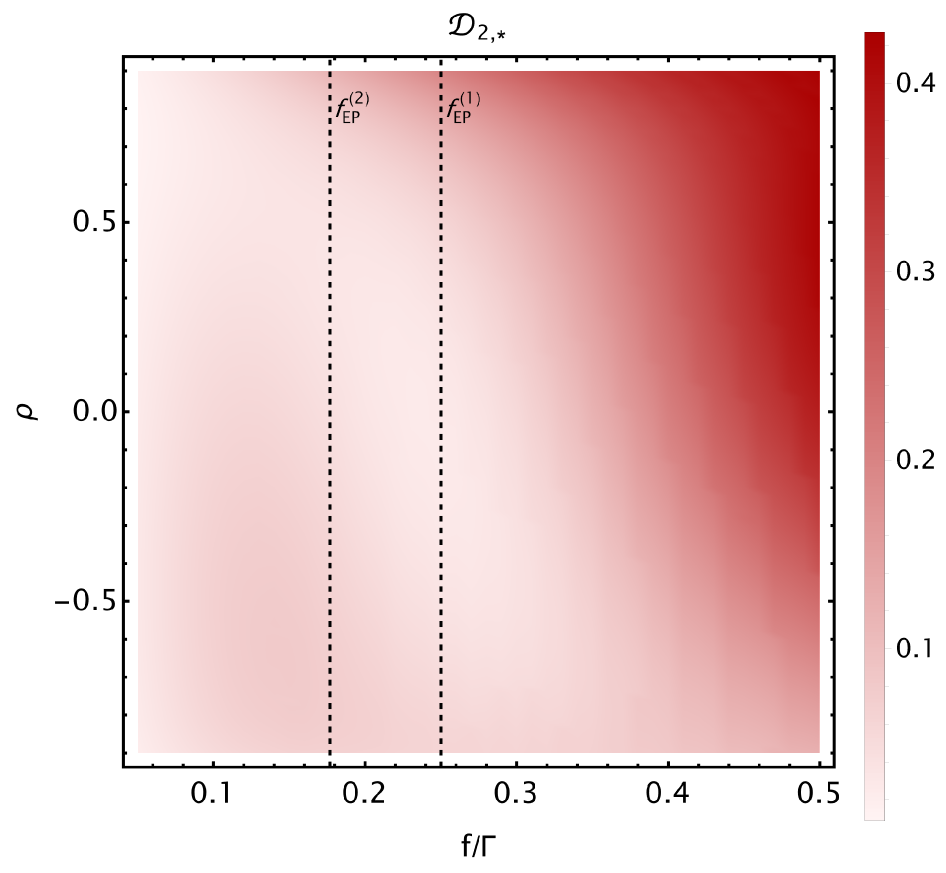}\\[-1mm]
\textbf{(a)}
\end{minipage}
\hfill
\begin{minipage}{0.48\textwidth}
\centering
\includegraphics[width=\linewidth]{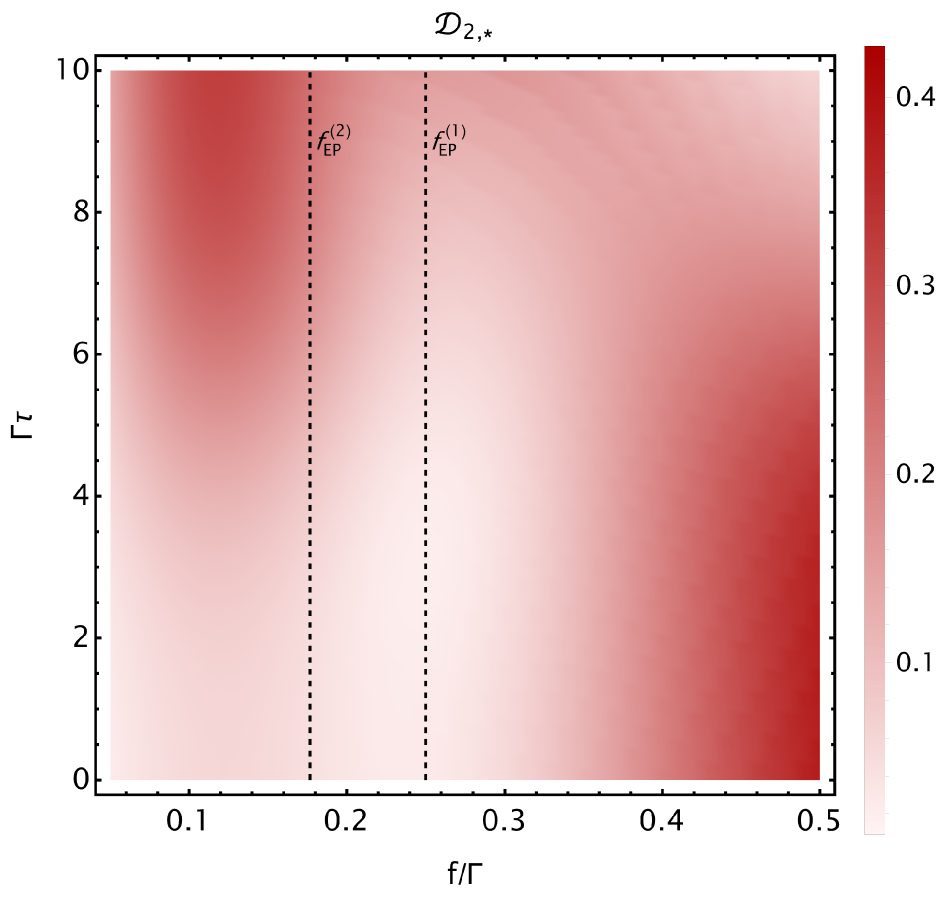}\\[-1mm]
\textbf{(b)}
\end{minipage}
\caption{Mean squared deviation from the fully charged battery state for temporally engineered two-photon pulses for $\Gamma T=2$. The color scale shows $\mathcal D_{2,*}=\langle(2-n_B)^2\rangle$ evaluated at the time $t_2^*$ at which $P_2(t)$ reaches its maximum. (a) Correlated Gaussian pulse as a function of the normalized TLS--battery coupling $f/\Gamma$ and the temporal-correlation coefficient $\rho$. (b) Symmetrized delayed Gaussian pulse as a function of $f/\Gamma$ and the dimensionless delay $\Gamma\tau$. The vertical dashed lines indicate the two exceptional points $f_{\rm EP}^{(2)}=\gamma/(2\sqrt{2})$ and $f_{\rm EP}^{(1)}=\gamma/2$. Lower values indicate a battery state closer to the fully charged state $|2\rangle_B$, and $\mathcal D_{2,*}=0$ occurs only for perfect charging, $P_2=1$.}
\label{fig:deviation_control}
\end{figure}

Figure~\ref{fig:deviation_control} presents $\mathcal D_{2,*}$ over the full pulse-control parameter space. Panel \ref{fig:deviation_control}(a) shows $\mathcal D_{2,*}(f/\Gamma,\rho)$ for correlated Gaussian pulses, while panel \ref{fig:deviation_control}(b) shows $\mathcal D_{2,*}(f/\Gamma,\Gamma\tau)$ for delayed ones. The vertical dashed lines mark the two exceptional points separating the three dynamical regimes.

The deviation maps should be interpreted together with the corresponding $P_2^{\max}$ maps in Figs.~\ref{fig:rho_gaussian} and \ref{fig:delayed_gaussian}. Regions of large $P_2^{\max}$ necessarily tend to have small $\mathcal D_{2,*}$, while the latter additionally distinguishes whether the remaining charging error is dominated by the empty-battery sector $P_0$, which carries weight four, or by the singly charged sector $P_1$, which carries weight one. Therefore, $\mathcal D_{2,*}$ provides a direct and unambiguous measure of the charging efficiency by showing the distance from the desired fully charged battery state.

\section{Discussion and Conclusions}
We have developed a general theory for charging a harmonic-oscillator quantum battery with an incoming two-photon wave packet. By reducing the problem to conditional single-photon and two-excitation response functions, we obtained a compact expression for the full charging amplitude. This expression shows that charging by a two-photon quantum pulse is governed by an intrinsically sequential temporal response: the first photon must partially transfer one excitation into the HO battery, before the second photon starts the charging process. The optimal incident state is therefore obtained by matching the joint temporal mode of the photon pair to the ordered response of the charger--battery system. This also explains why simultaneous photon arrival is generally suboptimal: the second excitation cannot be transferred efficiently before the first one has developed.

The internal dynamics of this two-step process are governed by two second-order exceptional points of the effective non-Hermitian amplitude generators. They separate overdamped, mixed, and underdamped charging regimes. In the mixed regime, the second-excitation response is already oscillatory while the first-excitation response remains overdamped. Above both exceptional points, both responses oscillate and change sign, so efficient charging requires matching not only the photon separation but also the time-dependent sign of the temporal amplitude. Applying this theory to Gaussian two-photon pulses, we found that temporal engineering of the photon pair can either enhance or suppress charging. Positive correlations concentrate the photon-pair probability near simultaneous interaction, where the sequential response is weak. In contrast, moderate negative correlations and finite delay can enhance the full-charging probability by matching the photon interaction-time separation to the internal TLS--battery response. Very large separations are again unfavorable because the first excitation decays or evolves away from the configuration required for efficient interaction of the second photon.

These results demonstrate that the temporal structure of a two-photon pulse provides an additional control parameter for quantum battery charging beyond pulse bandwidth and TLS--battery coupling strength. The arrival-time distribution and correlation orientation can both influence the transfer of the two available excitations into the battery. Temporally engineered few-photon states can therefore be used to optimize charging even when the total incident energy is fixed.

The temporal-mode analysis further shows that the degree of correlation alone is not a sufficient descriptor of charging performance. Correlated Gaussian states with coefficients $+\rho$ and $-\rho$ have the same purity and Schmidt number, but their joint temporal distributions are oriented differently. Negative correlations can align the photon-pair amplitude with the finite relative times selected by the ordered charging kernel, whereas positive correlations favor simultaneous interaction with TLS-HO. Hence, the dynamical value of temporal-mode entanglement is determined by its detailed mode structure and its matching to the system response, rather than by its magnitude alone.

The mean-squared target-number deviation $\mathcal D_{2,*}=\langle(2-n_B)^2\rangle_{t=t_2^*}$ complements the full-charging probability by quantifying the distance from the desired two-excitation battery state. Unlike the battery-energy variance, it vanishes only for perfect charging and therefore gives an unambiguous measure of residual charging error across the correlation- and delay-controlled protocols. 

The formal optimum establishes a broader response-matching principle: at a prescribed target time, the joint temporal wave function should reproduce the complex-conjugated ordered product of the response functions governing the first and second excitation transfers. The correlated and delayed Gaussian protocols considered here provide experimentally achievable approximations to this generally non-Gaussian optimal two-photon state.  Their charging performance is therefore determined by how closely they reproduce the temporal structure of the exact response-matched state.

Our results also complement continuously driven continuous-variable battery models~\cite{DowningUkhtary2025}. Such models are naturally suited to optimizing the amount and ergotropy of energy accumulated in a bosonic mode and can populate excitation sectors above two quanta. The present setting instead uses a fixed two-photon energy budget and resolves whether both quanta are stored in the prescribed state $|2\rangle_B$. Consequently, $\eta_E=E_B/(2\omega_b)$ quantifies energy-transfer efficiency, while $P_2$ and $\mathcal D_2$ provide additional state-selective information that is not contained in the mean energy alone. This distinction identifies temporal-mode engineering as a resource for controlled few-photon energy transfer rather than for unbounded energy accumulation.

In conclusion, we have developed an exact sequential-response theory for a harmonic-oscillator quantum battery charged by propagating two-photon light with a TLS acting as the charger. The results identify the response-matched joint temporal mode, clarify the role of the two effective exceptional points, and show that temporal anti-correlation and finite delay can improve charging within experimentally accessible Gaussian pulse families. 
In the ideal single-channel model, perfect charging is possible, while uncontrolled radiative loss imposes a branching-ratio-dependent upper bound. Photon temporal-correlation orientation, and time delay therefore provide control parameters beyond the coupling strength alone. Joint temporal-mode engineering offers a systematic route toward high-fidelity, target-state-selective energy transfer in microscopic quantum batteries.

\appendix
\section{Exact time-domain dynamics}
\label{app:time_dynamics}
This appendix gives the detailed solution of the time dynamics in the two-excitation manifold. Starting from the normal-ordered form of the Schr{\"o}dinger equation, we first derive the conditional single-photon response and then solve the coupled one- and two-excitation sectors, including the outgoing one- and two-photon amplitudes.

\subsection{Conditional single-photon response}
The input term in the normal-ordered Schr{\"o}dinger equation contains $-\sqrt{\Gamma}\sigma^+U(t)a(t)|\Psi(-\infty)\rangle$. We first evaluate the action of $a(t)$ on the symmetric two-photon input state $|2_{\phi_0}\rangle$. Using $[a(t),a^\dagger(\tau)]=\delta(t-\tau)$, we obtain
\begin{equation}
a(t)|\Psi(-\infty)\rangle = a(t)|2_{\phi_0}\rangle=\sqrt{2}\int du\,\phi_0(t,u)a^\dagger(u)|0\rangle .
\end{equation}
Thus, after the annihilation of one photon at time $t$, the remaining field is an effective single-photon with a generally unnormalized temporal wave packet $\xi_t(u)=\sqrt{2}\phi_0(u,t)$. 
For each fixed value of $t$, the subsequent evolution is exactly the single-photon charging problem with the replacement $\xi(u)\rightarrow \xi_t(u)$. 
\begin{equation}
\begin{aligned}
U(s)\left[|g\rangle|0\rangle_B\int du\,\xi_t(u)a^\dagger(u)|0\rangle_P\right]&=\alpha_0^b(s|t)\,|e\rangle|0\rangle_B|0\rangle_P+\alpha_1^b(s|t)\,|g\rangle|1\rangle_B|0\rangle_P\\
&+\int d\tau\,\phi^b_g(s;\tau|t)|g\rangle|0\rangle_Ba^\dagger(\tau)|0\rangle_P .
\end{aligned}
\end{equation}
We denote the corresponding conditional single-photon amplitudes by $\alpha_0^b(s|t)$, $\alpha_1^b(s|t)$, and $\phi_g^b(s;\tau|t)$. The vertical bar indicates that the effective single-photon evolution is conditioned on the annihilation of one photon from the initial two-photon state at time $t$. The amplitude $\alpha_0^b(s|t)$ corresponds to the state $|e\rangle|0\rangle_B|0\rangle_P$, while $\alpha_1^b(s|t)$ corresponds to $|g\rangle|1\rangle_B|0\rangle_P$. Finally, $\phi_g^b(s;\tau|t)$ is the temporal amplitude for the remaining photon to occupy the probe-field mode at time $\tau$, with the TLS and battery in their ground states. These amplitudes satisfy the same single-photon equations as in the one-photon battery problem, 
\begin{align} \label{ode1n}
\frac{d}{ds}\alpha_0^b(s|t)&= -\gamma \alpha_0^b(s|t)-if\alpha_1^b(s|t)-\sqrt{\Gamma}\xi_t(s),\\ \label{ode2n}
\frac{d}{ds}\alpha_1^b(s|t)&=-if\alpha_0^b(s|t),\\ \label{ode3n}
\frac{\partial}{\partial s} \phi^b_g(s,\tau|t)  &= \sqrt{\Gamma}  \delta(s-\tau) \, \alpha_0^b(s|t) 
\end{align}
with the initial conditions $\alpha_0^b(-\infty|t)=0,~\alpha_1^b(-\infty|t)=0$. The homogeneous part of Eqs.~\eqref{ode1n} and \eqref{ode2n} is generated by the single-excitation matrix as below, 
\begin{equation}
M_1=
\begin{pmatrix}
-\gamma & -if\\
-if & 0
\end{pmatrix}.
\end{equation}
The corresponding retarded propagator can be written as
\begin{equation}
R_1(t) =\Theta(t)e^{M_1t} =\Theta(t)
\begin{pmatrix}
\dot{G}_1(t) & -ifG_1(t)\\[4pt]
-ifG_1(t) & \dot{G}_1(t)+\gamma G_1(t)
\end{pmatrix}.
\label{R1_matrix}
\end{equation}
where the single-excitation response $G_1(t)$ is defined as,
\begin{equation}\label{G1_def}
G_1(t)= \frac{e^{r_+t}-e^{r_-t}}{r_+-r_-},\qquad r_\pm=\frac{-\gamma\pm\sqrt{\gamma^2-4f^2}}{2}.
\end{equation}
Thus, the general solution can be written as
\begin{align}
\alpha_0^b(s|t)&=-\sqrt{2\Gamma}\int_{-\infty}^s \dot{G}_1(s-u)\,\phi_0(u,t)\,du, \\ 
\alpha_1^b(s|t)&= i f \sqrt{2\Gamma}\int_{-\infty}^s G_1(s-u)\,\phi_0(u,t)\,du,\\
\phi^b_{g}(s,\tau|t)  &= \sqrt{2}\phi_0(\tau,t) + \sqrt{\Gamma}  \Theta(s-\tau) \alpha_0^b(\tau|t)
\end{align}
In particular, the equal-time value required in the two-photon charging equation is
\begin{equation}
\alpha_1^b(t|t)=if\sqrt{2\Gamma}\int_{-\infty}^{t}du\,G_1(t-u)\phi_0(u,t).
\end{equation}

\subsection{Two-excitation response and full charging amplitude}
Therefore, the normal-ordered equations  become
\begin{align}
\label{eq:alphae_ode}
\dot{\alpha}_e(t)&=-\gamma \alpha_e(t)-i\sqrt{2}f\alpha_g(t)-\sqrt{\Gamma}\alpha_1^b(t|t),\\
\label{eq:alphag_ode}
\dot{\alpha}_{g}(t)&=-i\sqrt{2}f\alpha_{e}(t),\\
\label{eq:chie_ode}
\partial_t\chi_{e}(t;\tau)&=-if\chi_{g}(t;\tau)-\gamma \chi_{e}(t;\tau)-\sqrt{\Gamma}\phi^b_g(t,\tau|t),\\
\label{eq:chig_ode}
\partial_t\chi_{g}(t;\tau)&=-if\chi_{e}(t;\tau)+\sqrt{\Gamma}\alpha_{e}(t)\delta(t-\tau),\\ 
\label{eq:phi_ode}
\partial_t\phi(t;\tau_1,\tau_2)&=\sqrt{\frac{\Gamma}{2}}\left[\delta(t-\tau_1)\chi_{e}(t;\tau_2)+\delta(t-\tau_2)\chi_{e}(t;\tau_1)\right].
\end{align}
The equations \eqref{eq:alphae_ode} and \eqref{eq:alphag_ode} can be written in vector form as
\begin{equation}
\frac{d}{dt}
\begin{pmatrix}
\alpha_e(t)\\
\alpha_g(t)
\end{pmatrix}
=M_2
\begin{pmatrix}
\alpha_e(t)\\
\alpha_g(t)
\end{pmatrix}
+
\begin{pmatrix}
-\sqrt{\Gamma}\alpha_1^b(t|t)\\
0
\end{pmatrix},
\label{two_excitation_vector_eq}
\end{equation}
with
\begin{equation}
M_2=
\begin{pmatrix}
-\gamma & -i\sqrt{2}f\\
-i\sqrt{2}f & 0
\end{pmatrix}.
\label{M2_def}
\end{equation}
The corresponding retarded propagator is
\begin{equation}
R_2(t)=\Theta(t)e^{M_2t}=\Theta(t)
\begin{pmatrix}
\dot{G}_2(t) & -i\sqrt{2}f\,G_2(t)\\[4pt]
-i\sqrt{2}f\,G_2(t) & \dot{G}_2(t)+\gamma G_2(t)
\end{pmatrix}.
\end{equation}
where the  two-excitation response function is defined as
\begin{equation} \label{G2_def}
G_2(t)=\frac{e^{q_+t}-e^{q_-t}}{q_+-q_-}, \qquad q_\pm=\frac{-\gamma\pm\sqrt{\gamma^2-8f^2}}{2}.
\end{equation}
Using the retarded solution of Eq.~\eqref{two_excitation_vector_eq}, we find
\begin{equation}
\alpha_g(t)=i\sqrt{2\Gamma}f\int_{-\infty}^{t}ds\,G_2(t-s)\alpha_1^b(s|s).
\label{alphag_before_substitution}
\end{equation}
Finally, we obtain the full two-excitation charging amplitude
\begin{equation}
\alpha_g(t)=-2\Gamma f^2\int_{-\infty}^{t}ds\int_{-\infty}^{s}du\, G_2(t-s)G_1(s-u)\phi_0(u,s).
\label{alphag_final_sym}
\end{equation}
This expression has a transparent sequential interpretation: the first photon interacts at time $u$, the second photon interacts at time $s$, and the system is observed at time $t$, with $u<s<t$. The function $G_1(s-u)$ describes the response associated with the first battery excitation, while $G_2(t-s)$ describes the response associated with the second battery excitation. The amplitude $\alpha_e(t)$ is obtained in the same way:
\begin{equation}
\alpha_e(t)=-\sqrt{\Gamma}\int_{-\infty}^{t}ds\,\dot{G}_2(t-s)\alpha_1^b(s|s),
\end{equation}
and therefore
\begin{equation}
\alpha_e(t)=-i\sqrt{2} \, \Gamma f\int_{-\infty}^{t}ds\int_{-\infty}^{s}du\,\dot{G}_2(t-s)G_1(s-u)\phi_0(u,s).
\label{alphae_final}
\end{equation}

\subsection{One-photon output sectors}
We now solve for the one-photon amplitudes $\chi_e(t;\tau)$ and $\chi_g(t;\tau)$. These amplitudes obey
\begin{equation}
\partial_t\chi_e(t;\tau)=-\gamma\chi_e(t;\tau)-if\chi_g(t;\tau)-\sqrt{\Gamma}\phi_g^b(t,\tau|t),
\label{chie_eq}
\end{equation}
\begin{equation}
\partial_t\chi_g(t;\tau)=-if\chi_e(t;\tau)+\sqrt{\Gamma}\alpha_e(t)\delta(t-\tau).
\label{chig_eq}
\end{equation}
Here $\phi_g^b(t,\tau|t)$ is the conditional single-photon output amplitude obtained from the one-photon problem:
\begin{equation}
\phi_g^b(s,\tau|t)=\xi_t(\tau)+\sqrt{\Gamma}\Theta(s-\tau)\alpha_0^b(\tau|t).
\label{phigb_def_recall}
\end{equation}
Therefore, for the source term in Eq.~\eqref{chie_eq}, we need
\begin{equation}
\phi_g^b(s,\tau|s)=\sqrt{2}\phi_0(s,\tau)+\sqrt{\Gamma}\Theta(s-\tau)\alpha_0^b(\tau|s).
\label{phigb_equal_condition}
\end{equation}

The homogeneous part of Eqs.~\eqref{chie_eq} and \eqref{chig_eq} is generated by the same single-excitation matrix that appeared in the conditional one-photon problem. Therefore, no new scalar Green function is required. We use the same single-excitation response $G_1(t)$ defined in Eq.~\eqref{G1_def}. The retarded solutions are then
\begin{equation}
\chi_e(t;\tau)=-\sqrt{\Gamma}\int_{-\infty}^{t}ds\,\dot{G}_1(t-s)\phi_g^b(s,\tau|s)-if\sqrt{\Gamma}\Theta(t-\tau)G_1(t-\tau)\alpha_e(\tau).
\label{chie_solution}
\end{equation}
Similarly,
\begin{equation}
\chi_g(t;\tau)=if\sqrt{\Gamma}\int_{-\infty}^{t}ds\,G_1(t-s)\phi_g^b(s,\tau|s)+\sqrt{\Gamma}\Theta(t-\tau)\left[\dot{G}_1(t-\tau)+\gamma G_1(t-\tau) \right] \alpha_e(\tau).
\label{chig_solution}
\end{equation}
The first term in each expression is driven by the conditional single-photon output field, while the second term comes from spontaneous emission from the two-excitation amplitude $\alpha_e(t)$ into the probe channel at time $t=\tau$.

\subsection{Two-photon output amplitude}
The two-photon output amplitude satisfies
\begin{equation}
\partial_t\phi(t;\tau_1,\tau_2)=\sqrt{\frac{\Gamma}{2}}\left[\delta(t-\tau_1)\chi_e(t;\tau_2)+\delta(t-\tau_2)\chi_e(t;\tau_1)\right].
\label{phi_eq}
\end{equation}
Integrating this equation from $-\infty$ to $t$, and using the initial condition
\begin{equation}
\phi(-\infty;\tau_1,\tau_2)=\phi_0(\tau_1,\tau_2),
\end{equation}
we obtain
\begin{equation}
\phi(t;\tau_1,\tau_2)=\phi_0(\tau_1,\tau_2)+\sqrt{\frac{\Gamma}{2}}\left[\Theta(t-\tau_1)\chi_e(\tau_1;\tau_2)+\Theta(t-\tau_2)\chi_e(\tau_2;\tau_1)\right].
\label{phi_solution}
\end{equation}
This expression is explicitly symmetric under $\tau_1\leftrightarrow\tau_2$, as required for a two-photon bosonic wave packet.

\section{Derivation of the optimal two-photon mode}
\label{app:optimal_two_photon}
At a fixed target time $t_f$, introduce the ordered-domain inner product
\begin{align}
\langle F,H\rangle_{\rm ord}=\int_{-\infty}^{t_f}ds
\int_{-\infty}^{s}du\,F^*(u,s)H(u,s).
\end{align}
Equation~\eqref{eq:alpha_kernel} becomes
\begin{align}
\alpha_g(t_f)=-2\Gamma f^2\left\langle
K_{t_f}^*,\phi_0\right\rangle_{\rm ord}.
\end{align}
For a normalized symmetric two-photon amplitude,
\begin{align}
\int dt_1dt_2\,|\phi_0(t_1,t_2)|^2=1,
\qquad
\phi_0(t_1,t_2)=\phi_0(t_2,t_1),
\end{align}
and, neglecting the measure-zero diagonal $t_1=t_2$,
\begin{align}
\int_{u<s}du\,ds\,|\phi_0(u,s)|^2=\frac{1}{2}.
\label{eq:half_norm}
\end{align}
Cauchy--Schwarz therefore gives
\begin{align}
|\alpha_g(t_f)|^2 &\le 4\Gamma^2f^4
\left[
\int_{u<s}du\,ds\,|K_{t_f}(u,s)|^2
\right]
\left[
\int_{u<s}du\,ds\,|\phi_0(u,s)|^2
\right]
\nonumber\\
&=
2\Gamma^2f^4\,\mathcal N_2(t_f),
\end{align}
which proves Eq.~\eqref{eq:P2_bound}.

The normalization integral $\mathcal N_2(t_f)$ can be evaluated analytically. Introducing the nonnegative time differences
\begin{equation}
x=t_f-s,\qquad y=s-u,
\end{equation}
maps the ordered domain $u<s<t_f$ onto $x,y\in[0,\infty)$ and factorizes Eq.~\eqref{eq:N2} as
\begin{equation}
\mathcal N_2=\left[\int_0^\infty dx\,|G_2(x)|^2\right]
\left[\int_0^\infty dy\,|G_1(y)|^2\right].
\label{eq:N2_factorized}
\end{equation}
The response function in excitation sector $j=1,2$ obeys
\begin{equation}
\ddot G_j(t)+\gamma\dot G_j(t)+jf^2G_j(t)=0,
\qquad
G_j(0)=0,\qquad \dot G_j(0)=1.
\label{eq:Gj_ode}
\end{equation}
Defining
\begin{equation}
I_j=\int_0^\infty dt\,|G_j(t)|^2,
\qquad
J_j=\int_0^\infty dt\,|\dot G_j(t)|^2,
\end{equation}
multiplication of Eq.~\eqref{eq:Gj_ode} by $G_j(t)$, followed by integration by parts, gives
\begin{equation}
J_j=jf^2 I_j.
\label{eq:Ij_Jj_relation}
\end{equation}
Similarly, multiplication by $\dot G_j(t)$ gives
\begin{equation}
-\frac{1}{2}+\gamma J_j=0,
\end{equation}
where the boundary conditions $G_j(\infty)=\dot G_j(\infty)=0$ have been used. Hence,
\begin{equation}
\int_0^\infty dt\,|G_j(t)|^2 = \frac{1}{2\gamma jf^2}.
\label{eq:Gj_norm}
\end{equation}
This result is valid in the overdamped, exceptional-point, and
underdamped regimes. Equation~\eqref{eq:N2_factorized} therefore yields
\begin{equation}
\mathcal N_2(t_f)=\frac{1}{8\gamma^2f^4}.
\label{eq:N2_analytic}
\end{equation}
The result is independent of $t_f$, because changing the target time only translates the optimal temporal mode. Substitution into Eq.~\eqref{eq:P2_bound} gives
\begin{equation}
P_{2,\mathrm{opt}} = 2\Gamma^2f^4\mathcal N_2 = \frac{\Gamma^2}{4\gamma^2},
\end{equation}
which gives Eq.~\eqref{eq:P2_ultimate} in the main text.

Equality is obtained when the wave packet restricted to the ordered region is proportional to the conjugated kernel:
\begin{align}
\phi_{\rm opt}(u,s;t_f)=\frac{K_{t_f}^*(u,s)}
{\sqrt{2\mathcal N_2(t_f)}}, \qquad u<s<t_f.
\label{eq:phi_ordered}
\end{align}
Extending Eq.~\eqref{eq:phi_ordered} to the opposite ordering by bosonic symmetry yields
\begin{align}
\phi_{\rm opt}(t_1,t_2;t_f)=\frac{\mathcal K_{t_f}^*(t_1,t_2)
+\mathcal K_{t_f}^*(t_2,t_1)} {\sqrt{2\mathcal N_2(t_f)}}.
\end{align}
The supports of the two ordered terms are disjoint except on the diagonal, so the normalization follows immediately:
\begin{align}
\int dt_1dt_2\,|\phi_{\rm opt}(t_1,t_2;t_f)|^2
=\frac{2\mathcal N_2(t_f)}{2\mathcal N_2(t_f)} =1.
\end{align}
The associated maximal probability at the prescribed target time is
\begin{align}
P_{2,\rm opt}(t_f)=2\Gamma^2f^4\,\mathcal N_2(t_f).
\label{eq:P2opt_tf}
\end{align}
A global optimum over charging time is then obtained from
\begin{align}
P_{2,\rm opt}^{\max}=\max_{t_f}P_{2,\rm opt}(t_f).
\label{eq:P2opt_global}
\end{align}

\section*{Acknowledgment}
E.D. thanks Dr. Charles Andrew Downing for an early discussion that motivated the initiation of this two-photon extension. E.D and M.K. acknowledge funding by the Higher Education and Science Committee of Armenia under Grant No. 22IRF-06.

\bibliographystyle{apsrev4-2}
\bibliography{references}

\begin{thebibliography}{21}%
\makeatletter
\providecommand \@ifxundefined [1]{%
 \@ifx{#1\undefined}
}%
\providecommand \@ifnum [1]{%
 \ifnum #1\expandafter \@firstoftwo
 \else \expandafter \@secondoftwo
 \fi
}%
\providecommand \@ifx [1]{%
 \ifx #1\expandafter \@firstoftwo
 \else \expandafter \@secondoftwo
 \fi
}%
\providecommand \natexlab [1]{#1}%
\providecommand \enquote  [1]{``#1''}%
\providecommand \bibnamefont  [1]{#1}%
\providecommand \bibfnamefont [1]{#1}%
\providecommand \citenamefont [1]{#1}%
\providecommand \href@noop [0]{\@secondoftwo}%
\providecommand \href [0]{\begingroup \@sanitize@url \@href}%
\providecommand \@href[1]{\@@startlink{#1}\@@href}%
\providecommand \@@href[1]{\endgroup#1\@@endlink}%
\providecommand \@sanitize@url [0]{\catcode `\\12\catcode `\$12\catcode
  `\&12\catcode `\#12\catcode `\^12\catcode `\_12\catcode `\%12\relax}%
\providecommand \@@startlink[1]{}%
\providecommand \@@endlink[0]{}%
\providecommand \url  [0]{\begingroup\@sanitize@url \@url }%
\providecommand \@url [1]{\endgroup\@href {#1}{\urlprefix }}%
\providecommand \urlprefix  [0]{URL }%
\providecommand \Eprint [0]{\href }%
\providecommand \doibase [0]{https://doi.org/}%
\providecommand \selectlanguage [0]{\@gobble}%
\providecommand \bibinfo  [0]{\@secondoftwo}%
\providecommand \bibfield  [0]{\@secondoftwo}%
\providecommand \translation [1]{[#1]}%
\providecommand \BibitemOpen [0]{}%
\providecommand \bibitemStop [0]{}%
\providecommand \bibitemNoStop [0]{.\EOS\space}%
\providecommand \EOS [0]{\spacefactor3000\relax}%
\providecommand \BibitemShut  [1]{\csname bibitem#1\endcsname}%
\let\auto@bib@innerbib\@empty
\bibitem [{\citenamefont {Alicki}\ and\ \citenamefont
  {Fannes}(2013)}]{AlickiFannes2013}%
  \BibitemOpen
  \bibfield  {author} {\bibinfo {author} {\bibfnamefont {R.}~\bibnamefont
  {Alicki}}\ and\ \bibinfo {author} {\bibfnamefont {M.}~\bibnamefont
  {Fannes}},\ }\href {https://doi.org/10.1103/PhysRevE.87.042123} {\bibfield
  {journal} {\bibinfo  {journal} {Phys. Rev. E}\ }\textbf {\bibinfo {volume}
  {87}},\ \bibinfo {pages} {042123} (\bibinfo {year} {2013})}\BibitemShut
  {NoStop}%
\bibitem [{\citenamefont {Binder}\ \emph {et~al.}(2015)\citenamefont {Binder},
  \citenamefont {Vinjanampathy}, \citenamefont {Modi},\ and\ \citenamefont
  {Goold}}]{Binder2015}%
  \BibitemOpen
  \bibfield  {author} {\bibinfo {author} {\bibfnamefont {F.~C.}\ \bibnamefont
  {Binder}}, \bibinfo {author} {\bibfnamefont {S.}~\bibnamefont
  {Vinjanampathy}}, \bibinfo {author} {\bibfnamefont {K.}~\bibnamefont
  {Modi}},\ and\ \bibinfo {author} {\bibfnamefont {J.}~\bibnamefont {Goold}},\
  }\href {https://doi.org/10.1088/1367-2630/17/7/075015} {\bibfield  {journal}
  {\bibinfo  {journal} {New J. Phys.}\ }\textbf {\bibinfo {volume} {17}},\
  \bibinfo {pages} {075015} (\bibinfo {year} {2015})}\BibitemShut {NoStop}%
\bibitem [{\citenamefont {Campaioli}\ \emph {et~al.}(2017)\citenamefont
  {Campaioli}, \citenamefont {Pollock}, \citenamefont {Binder}, \citenamefont
  {C{\'e}leri}, \citenamefont {Goold}, \citenamefont {Vinjanampathy},\ and\
  \citenamefont {Modi}}]{Campaioli2017}%
  \BibitemOpen
  \bibfield  {author} {\bibinfo {author} {\bibfnamefont {F.}~\bibnamefont
  {Campaioli}}, \bibinfo {author} {\bibfnamefont {F.~A.}\ \bibnamefont
  {Pollock}}, \bibinfo {author} {\bibfnamefont {F.~C.}\ \bibnamefont {Binder}},
  \bibinfo {author} {\bibfnamefont {L.~C.}\ \bibnamefont {C{\'e}leri}},
  \bibinfo {author} {\bibfnamefont {J.}~\bibnamefont {Goold}}, \bibinfo
  {author} {\bibfnamefont {S.}~\bibnamefont {Vinjanampathy}},\ and\ \bibinfo
  {author} {\bibfnamefont {K.}~\bibnamefont {Modi}},\ }\href
  {https://doi.org/10.1103/PhysRevLett.118.150601} {\bibfield  {journal}
  {\bibinfo  {journal} {Phys. Rev. Lett.}\ }\textbf {\bibinfo {volume} {118}},\
  \bibinfo {pages} {150601} (\bibinfo {year} {2017})}\BibitemShut {NoStop}%
\bibitem [{\citenamefont {Ferraro}\ \emph {et~al.}(2018)\citenamefont
  {Ferraro}, \citenamefont {Campisi}, \citenamefont {Andolina}, \citenamefont
  {Pellegrini},\ and\ \citenamefont {Polini}}]{Ferraro2018}%
  \BibitemOpen
  \bibfield  {author} {\bibinfo {author} {\bibfnamefont {D.}~\bibnamefont
  {Ferraro}}, \bibinfo {author} {\bibfnamefont {M.}~\bibnamefont {Campisi}},
  \bibinfo {author} {\bibfnamefont {G.~M.}\ \bibnamefont {Andolina}}, \bibinfo
  {author} {\bibfnamefont {V.}~\bibnamefont {Pellegrini}},\ and\ \bibinfo
  {author} {\bibfnamefont {M.}~\bibnamefont {Polini}},\ }\href
  {https://doi.org/10.1103/PhysRevLett.120.117702} {\bibfield  {journal}
  {\bibinfo  {journal} {Phys. Rev. Lett.}\ }\textbf {\bibinfo {volume} {120}},\
  \bibinfo {pages} {117702} (\bibinfo {year} {2018})}\BibitemShut {NoStop}%
\bibitem [{\citenamefont {Campaioli}\ \emph {et~al.}(2024)\citenamefont
  {Campaioli}, \citenamefont {Gherardini}, \citenamefont {Quach}, \citenamefont
  {Polini},\ and\ \citenamefont {Andolina}}]{CampaioliRMP2024}%
  \BibitemOpen
  \bibfield  {author} {\bibinfo {author} {\bibfnamefont {F.}~\bibnamefont
  {Campaioli}}, \bibinfo {author} {\bibfnamefont {S.}~\bibnamefont
  {Gherardini}}, \bibinfo {author} {\bibfnamefont {J.~Q.}\ \bibnamefont
  {Quach}}, \bibinfo {author} {\bibfnamefont {M.}~\bibnamefont {Polini}},\ and\
  \bibinfo {author} {\bibfnamefont {G.~M.}\ \bibnamefont {Andolina}},\ }\href
  {https://doi.org/10.1103/RevModPhys.96.031001} {\bibfield  {journal}
  {\bibinfo  {journal} {Rev. Mod. Phys.}\ }\textbf {\bibinfo {volume} {96}},\
  \bibinfo {pages} {031001} (\bibinfo {year} {2024})}\BibitemShut {NoStop}%
\bibitem [{\citenamefont {Downing}\ and\ \citenamefont
  {Ukhtary}(2024)}]{DowningUkhtaryPLA2024}%
  \BibitemOpen
  \bibfield  {author} {\bibinfo {author} {\bibfnamefont {C.~A.}\ \bibnamefont
  {Downing}}\ and\ \bibinfo {author} {\bibfnamefont {M.~S.}\ \bibnamefont
  {Ukhtary}},\ }\href {https://doi.org/10.1016/j.physleta.2024.129693}
  {\bibfield  {journal} {\bibinfo  {journal} {Phys. Lett. A}\ }\textbf
  {\bibinfo {volume} {518}},\ \bibinfo {pages} {129693} (\bibinfo {year}
  {2024})}\BibitemShut {NoStop}%
\bibitem [{\citenamefont {Downing}\ and\ \citenamefont
  {Ukhtary}(2025)}]{DowningUkhtary2025}%
  \BibitemOpen
  \bibfield  {author} {\bibinfo {author} {\bibfnamefont {C.~A.}\ \bibnamefont
  {Downing}}\ and\ \bibinfo {author} {\bibfnamefont {M.~S.}\ \bibnamefont
  {Ukhtary}},\ }\href {https://doi.org/10.1103/73zl-yn4h} {\bibfield  {journal}
  {\bibinfo  {journal} {Phys. Rev. E}\ }\textbf {\bibinfo {volume} {112}},\
  \bibinfo {pages} {044143} (\bibinfo {year} {2025})}\BibitemShut {NoStop}%
\bibitem [{\citenamefont {Darsheshdar}\ and\ \citenamefont
  {Moniri}(2026)}]{DarsheshdarMoniri2026}%
  \BibitemOpen
  \bibfield  {author} {\bibinfo {author} {\bibfnamefont {E.}~\bibnamefont
  {Darsheshdar}}\ and\ \bibinfo {author} {\bibfnamefont {S.~M.}\ \bibnamefont
  {Moniri}},\ }\href {https://doi.org/10.1088/1361-6455/ae7eb7} {\bibfield
  {journal} {\bibinfo  {journal} {J. Phys. B: At. Mol. Opt. Phys.}\ }\textbf
  {\bibinfo {volume} {59}},\ \bibinfo {pages} {135501} (\bibinfo {year}
  {2026})}\BibitemShut {NoStop}%
\bibitem [{\citenamefont {Khanbekyan}\ \emph {et~al.}(2008)\citenamefont
  {Khanbekyan}, \citenamefont {Welsch}, \citenamefont {Di~Fidio},\ and\
  \citenamefont {Vogel}}]{Khanbekyan2008Pulse}%
  \BibitemOpen
  \bibfield  {author} {\bibinfo {author} {\bibfnamefont {M.}~\bibnamefont
  {Khanbekyan}}, \bibinfo {author} {\bibfnamefont {D.-G.}\ \bibnamefont
  {Welsch}}, \bibinfo {author} {\bibfnamefont {C.}~\bibnamefont {Di~Fidio}},\
  and\ \bibinfo {author} {\bibfnamefont {W.}~\bibnamefont {Vogel}},\ }\href
  {https://doi.org/10.1103/PhysRevA.78.013822} {\bibfield  {journal} {\bibinfo
  {journal} {Phys. Rev. A}\ }\textbf {\bibinfo {volume} {78}},\ \bibinfo
  {pages} {013822} (\bibinfo {year} {2008})}\BibitemShut {NoStop}%
\bibitem [{\citenamefont {Khanbekyan}\ and\ \citenamefont
  {Welsch}(2017)}]{Khanbekyan2017Controlled}%
  \BibitemOpen
  \bibfield  {author} {\bibinfo {author} {\bibfnamefont {M.}~\bibnamefont
  {Khanbekyan}}\ and\ \bibinfo {author} {\bibfnamefont {D.-G.}\ \bibnamefont
  {Welsch}},\ }\href {https://doi.org/10.1103/PhysRevA.95.013803} {\bibfield
  {journal} {\bibinfo  {journal} {Phys. Rev. A}\ }\textbf {\bibinfo {volume}
  {95}},\ \bibinfo {pages} {013803} (\bibinfo {year} {2017})}\BibitemShut
  {NoStop}%
\bibitem [{\citenamefont {Khanbekyan}(2018)}]{Khanbekyan2018TimeBin}%
  \BibitemOpen
  \bibfield  {author} {\bibinfo {author} {\bibfnamefont {M.}~\bibnamefont
  {Khanbekyan}},\ }\href {https://doi.org/10.3103/S1068337218040035} {\bibfield
   {journal} {\bibinfo  {journal} {J. Contemp. Phys.}\ }\textbf {\bibinfo
  {volume} {53}},\ \bibinfo {pages} {286} (\bibinfo {year} {2018})}\BibitemShut
  {NoStop}%
\bibitem [{\citenamefont {Albarelli}\ \emph {et~al.}(2023)\citenamefont
  {Albarelli}, \citenamefont {Bisketzi}, \citenamefont {Khan},\ and\
  \citenamefont {Datta}}]{Albarelli2023}%
  \BibitemOpen
  \bibfield  {author} {\bibinfo {author} {\bibfnamefont {F.}~\bibnamefont
  {Albarelli}}, \bibinfo {author} {\bibfnamefont {E.}~\bibnamefont {Bisketzi}},
  \bibinfo {author} {\bibfnamefont {A.}~\bibnamefont {Khan}},\ and\ \bibinfo
  {author} {\bibfnamefont {A.}~\bibnamefont {Datta}},\ }\href
  {https://doi.org/10.1103/PhysRevA.107.062601} {\bibfield  {journal} {\bibinfo
   {journal} {Phys. Rev. A}\ }\textbf {\bibinfo {volume} {107}},\ \bibinfo
  {pages} {062601} (\bibinfo {year} {2023})}\BibitemShut {NoStop}%
\bibitem [{\citenamefont {Darsheshdar}\ \emph {et~al.}(2024)\citenamefont
  {Darsheshdar}, \citenamefont {Khan}, \citenamefont {Albarelli},\ and\
  \citenamefont {Datta}}]{Darsheshdar2024Chirp}%
  \BibitemOpen
  \bibfield  {author} {\bibinfo {author} {\bibfnamefont {E.}~\bibnamefont
  {Darsheshdar}}, \bibinfo {author} {\bibfnamefont {A.}~\bibnamefont {Khan}},
  \bibinfo {author} {\bibfnamefont {F.}~\bibnamefont {Albarelli}},\ and\
  \bibinfo {author} {\bibfnamefont {A.}~\bibnamefont {Datta}},\ }\href
  {https://doi.org/10.1103/PhysRevA.110.043710} {\bibfield  {journal} {\bibinfo
   {journal} {Phys. Rev. A}\ }\textbf {\bibinfo {volume} {110}},\ \bibinfo
  {pages} {043710} (\bibinfo {year} {2024})}\BibitemShut {NoStop}%
\bibitem [{\citenamefont {Khan}\ \emph {et~al.}(2024)\citenamefont {Khan},
  \citenamefont {Albarelli},\ and\ \citenamefont {Datta}}]{Khan2024QST}%
  \BibitemOpen
  \bibfield  {author} {\bibinfo {author} {\bibfnamefont {A.}~\bibnamefont
  {Khan}}, \bibinfo {author} {\bibfnamefont {F.}~\bibnamefont {Albarelli}},\
  and\ \bibinfo {author} {\bibfnamefont {A.}~\bibnamefont {Datta}},\ }\href
  {https://doi.org/10.1088/2058-9565/ad331b} {\bibfield  {journal} {\bibinfo
  {journal} {Quantum Sci. Technol.}\ }\textbf {\bibinfo {volume} {9}},\
  \bibinfo {pages} {035004} (\bibinfo {year} {2024})}\BibitemShut {NoStop}%
\bibitem [{\citenamefont {Khan}\ \emph {et~al.}(2025)\citenamefont {Khan},
  \citenamefont {Albarelli},\ and\ \citenamefont {Datta}}]{Khan2025Tensor}%
  \BibitemOpen
  \bibfield  {author} {\bibinfo {author} {\bibfnamefont {A.}~\bibnamefont
  {Khan}}, \bibinfo {author} {\bibfnamefont {F.}~\bibnamefont {Albarelli}},\
  and\ \bibinfo {author} {\bibfnamefont {A.}~\bibnamefont {Datta}},\ }\href
  {https://doi.org/10.1103/ljh3-3l4j} {\bibfield  {journal} {\bibinfo
  {journal} {PRX Quantum}\ }\textbf {\bibinfo {volume} {6}},\ \bibinfo {pages}
  {040343} (\bibinfo {year} {2025})}\BibitemShut {NoStop}%
\bibitem [{\citenamefont {Das}\ \emph {et~al.}(2025)\citenamefont {Das},
  \citenamefont {Khan}, \citenamefont {Albarelli},\ and\ \citenamefont
  {Datta}}]{Das2025Vibrational}%
  \BibitemOpen
  \bibfield  {author} {\bibinfo {author} {\bibfnamefont {S.}~\bibnamefont
  {Das}}, \bibinfo {author} {\bibfnamefont {A.}~\bibnamefont {Khan}}, \bibinfo
  {author} {\bibfnamefont {F.}~\bibnamefont {Albarelli}},\ and\ \bibinfo
  {author} {\bibfnamefont {A.}~\bibnamefont {Datta}},\ }\href@noop {}
  {\bibfield  {journal} {\bibinfo  {journal} {arXiv preprint arXiv:2510.08386}\
  } (\bibinfo {year} {2025})}\BibitemShut {NoStop}%
\bibitem [{\citenamefont {Baragiola}\ \emph {et~al.}(2012)\citenamefont
  {Baragiola}, \citenamefont {Cook}, \citenamefont {Bra{\'n}czyk},\ and\
  \citenamefont {Combes}}]{Baragiola2012}%
  \BibitemOpen
  \bibfield  {author} {\bibinfo {author} {\bibfnamefont {B.~Q.}\ \bibnamefont
  {Baragiola}}, \bibinfo {author} {\bibfnamefont {R.~L.}\ \bibnamefont {Cook}},
  \bibinfo {author} {\bibfnamefont {A.~M.}\ \bibnamefont {Bra{\'n}czyk}},\ and\
  \bibinfo {author} {\bibfnamefont {J.}~\bibnamefont {Combes}},\ }\href
  {https://doi.org/10.1103/PhysRevA.86.013811} {\bibfield  {journal} {\bibinfo
  {journal} {Phys. Rev. A}\ }\textbf {\bibinfo {volume} {86}},\ \bibinfo
  {pages} {013811} (\bibinfo {year} {2012})}\BibitemShut {NoStop}%
\bibitem [{\citenamefont {Khanbekyan}\ and\ \citenamefont
  {Wiersig}(2020)}]{KhanbekyanWiersig2020}%
  \BibitemOpen
  \bibfield  {author} {\bibinfo {author} {\bibfnamefont {M.}~\bibnamefont
  {Khanbekyan}}\ and\ \bibinfo {author} {\bibfnamefont {J.}~\bibnamefont
  {Wiersig}},\ }\href {https://doi.org/10.1103/PhysRevResearch.2.023375}
  {\bibfield  {journal} {\bibinfo  {journal} {Phys. Rev. Research}\ }\textbf
  {\bibinfo {volume} {2}},\ \bibinfo {pages} {023375} (\bibinfo {year}
  {2020})}\BibitemShut {NoStop}%
\bibitem [{\citenamefont {Khanbekyan}(2023)}]{Khanbekyan2023EP}%
  \BibitemOpen
  \bibfield  {author} {\bibinfo {author} {\bibfnamefont {M.}~\bibnamefont
  {Khanbekyan}},\ }\href {https://doi.org/10.1103/PhysRevA.108.023710}
  {\bibfield  {journal} {\bibinfo  {journal} {Phys. Rev. A}\ }\textbf {\bibinfo
  {volume} {108}},\ \bibinfo {pages} {023375} (\bibinfo {year}
  {2023})}\BibitemShut {NoStop}%
\bibitem [{\citenamefont {Mikhailova}\ \emph {et~al.}(2008)\citenamefont
  {Mikhailova}, \citenamefont {Volkov},\ and\ \citenamefont
  {Fedorov}}]{Mikhailova2008}%
  \BibitemOpen
  \bibfield  {author} {\bibinfo {author} {\bibfnamefont {Y.~M.}\ \bibnamefont
  {Mikhailova}}, \bibinfo {author} {\bibfnamefont {P.~A.}\ \bibnamefont
  {Volkov}},\ and\ \bibinfo {author} {\bibfnamefont {M.~v.}\ \bibnamefont
  {Fedorov}},\ }\href {https://doi.org/10.1103/PhysRevA.78.062327} {\bibfield
  {journal} {\bibinfo  {journal} {Phys. Rev. A}\ }\textbf {\bibinfo {volume}
  {78}},\ \bibinfo {pages} {062327} (\bibinfo {year} {2008})}\BibitemShut
  {NoStop}%
\bibitem [{\citenamefont {Jeronimo-Moreno}\ and\ \citenamefont
  {U'Ren}(2009)}]{JeronimoMoreno2009}%
  \BibitemOpen
  \bibfield  {author} {\bibinfo {author} {\bibfnamefont {Y.}~\bibnamefont
  {Jeronimo-Moreno}}\ and\ \bibinfo {author} {\bibfnamefont {A.~B.}\
  \bibnamefont {U'Ren}},\ }\href {https://doi.org/10.1103/PhysRevA.79.033839}
  {\bibfield  {journal} {\bibinfo  {journal} {Phys. Rev. A}\ }\textbf {\bibinfo
  {volume} {79}},\ \bibinfo {pages} {033839} (\bibinfo {year}
  {2009})}\BibitemShut {NoStop}%
\end{thebibliography}%

\end{document}